\documentstyle[epsf,seceq,preprint]{jpsj}
\title
{Josephson-Vortex-Glass Transition in Strong Fields}
\author{Ryusuke Ikeda and Hiroto Adachi}
\inst{Department of Physics, Kyoto University, Kyoto 606-8502}
\recdate{March 29, 2000}

\abst{
A vortex-glass transition due to point disorder in layered 
superconductors is studied for the case with an applied field 
{\it parallel} to the layers. Our calculation of tilt responses 
indicates that, irrespective of the magnitude of the field, the 
resulting glass phase, Josephson-vortex-glass (JG), should have a 
transverse Meissner effect, as in a planar splayed glass phase, 
only for a tilt perpendicular to the layers. 
Further, focusing on the high field (and/or high anisotropy) region 
$B \sqrt{\Gamma} > \phi_0/d^2$, where $\Gamma$ is the mass anisotropy 
in the Lawrence-Doniach model, the JG transition line $T_{JG}(B)$ 
is shown to have a similar form to a $B$-$T$ line following from 
the {\it disorder-free} Lindemann criterion and to decrease with
increasing $B \sqrt{\Gamma}$, in marked contrast to the 
disorder-free melting line {\it insensitive to} $B \sqrt{\Gamma}$ 
in such the high field region. This $T_{JG}(B)$ line seems to have been 
recently observed in a.c. susceptibility and in-plane resistivity
measurements in BSCCO and qualitatively explains a field dependence 
at lower temperatures of previous BSCCO resistivity data showing the 
so-called in-plane Lorentz force-free behavior. 
}

\kword{Type II superconductors, vortex states, Josephson vortex, 
vortex-glass}
\begin{document}
\sloppy
\maketitle

\section{Introduction}

In bulk type II superconductors, different types of superconducting
glass phases with no Ohmic dissipation 
are expected, depending on types and configurations of
the static disorder (randomness), to be realized in a magnetic field. At
least in homogeneous materials (i.e., with no granularity), no glass
ordering is expected in zero field without transition to the Meissner 
phase because there is no origin of frustration. By contrast, in a 
nonzero field a spatial variation of $|\psi|$ due to field-induced
vortices makes a formation of a glass phase with vanishing resistivity 
rather favorable at a low enough temperature.\cite{RI1,GD,Natter} 
In this sense,
the superconducting glass phases are peculiar to the nonzero field 
case and hence, can be called the vortex-glass phases.\cite{FFH,RI2} 

So far, Ginzburg-Landau (GL) studies\cite{RI1,RI2} of the vortex-glass 
transitions in a layered superconductor were limited to the most
familiar  
case (${\bf B} \parallel {\bf c}$) with a field perpendicular to the
layers. No GL description of the case (${\bf B} \perp {\bf c}$) in a 
field parallel to the layers and with no disorder was available even
semiquantitatively until recently, and it was first performed in ref.5 
in an approximated but systematic manner focusing on the lowest Landau 
level (LLL) modes of the pair-field $\psi$ (superconducting order 
parameter). Since a systematic description of the glass {\it transitions}
in ${\bf B} \parallel {\bf c}$ is available at present based only on the
LLL approach, it is natural to, based on the results in ref.5, extend 
our study of a vortex-glass transition to ${\bf B} \perp {\bf c}$
case. It is our purpose in this paper to perform this.  

The present study was motivated in part by resistive data of high $T_c$ 
materials. First, the vanishing behaviors of Ohmic dissipation in
real YBCO materials in ${\bf B} \perp {\bf c}$ are
controversial.\cite{Kwok,Chara,Ishida,Nigel} Resistive data\cite{Kwok} 
of 90 K YBCO for a current perpendicular
to ${\bf B}$ and parallel to the layers have indicated 
a continuous thermodynamic transition, while the ${\bf c}$-axis
resistive ($\rho_c$) data in ref.7 have shown a first order transition 
as well as 
in ${\bf B} \parallel {\bf c}$. More recently, a resistive 
measurement\cite{Nigel}
in YBa$_2$Cu$_4$O$_8$ has shown an {\it enhancement} of a sharp 
vanishing, suggestive of a thermodynamic first 
order transition, of $\rho_c$ with
decreasing the tilt angle between the ${\bf B}$ direction and the CuO
planes. Such a vanishing, at least, of $\rho_c$ in ${\bf B} \perp {\bf
c}$ cannot be understood
without point disorder on the CuO layers.\cite{RI3} 
Further, since the vanishing of
resistivities in vortex systems is generally of a collective nature, 
a collective disorder effect near a freezing transition must appear 
irrespective of the current direction. Clearly, these 
observations\cite{Kwok,Chara,Nigel} need 
to be discussed in the context of a point-disorder-induced glass
transition. 

Second, to our knowledge, the issue of the so-called in-plane Lorentz 
force free
behavior\cite{Iye} in BSCCO in tesla range has not been resolved
sufficiently. In tesla range of this material satisfying $B \gg
\phi_0/(2 \pi \sqrt{\Gamma}d^2)$ (see $\S 2$), the resistive data are
quantitatively independent of 
the relative angle between ${\bf B}$ and the applied current, both of
which are applied in directions perpendicular to the ${\bf c}$ axis. 
This ``force free'' behavior was verified later in refs.11 and 12.  
To explain the data in ref.10, a (technically inevitable) deviation 
of the applied field from the CuO plane\cite{Kes} and the presence of
the so-called double-kink excitation\cite{Iye2} were argued in 
literatures. The former\cite{Kes} interpretation ascribing
the angular independence to a vortex flow of deviation-induced pancake 
vortices inevitably contradicts the tendency
seen in Fig.2 of ref.11 that, at higher temperatures (around $86 {\rm K}$) 
below $T_{c0}$, the field dependence of resistivity disappears 
with {\it increasing} $B$. Further, the picture, used in ref.12, 
of a vortex liquid confined within the interlayer spacings {\it with} 
double-kinks thermally-excited cannot\cite{Blatter} explain Ohmic 
dissipation seen\cite{Kad,Iye2} in the disordered phase. 
In ref.15, one of us has argued 
based on the anisotropic GL fluctuation theory that, with no static 
disorder, the in-plane Lorentz force free behavior is due merely to 
a strong enough
anisotropy and will simultaneously be accompanied by 
the absence of a $B$ dependence
of resistive data in the same $B$ range. Although a statement similar to 
this remains valid in the Lawrence-Doniach (LD) model 
with layer structure inducing
the intrinsic pinning effect (see $\S 2$) and explains the 
above-mentioned behavior ($B$-{\it independent} ``force free'' 
behavior) in ref.11 at higher temperatures, the resistive data in 
Fig.2 of refs.11 and 12 seem to show a $B$-{\it dependence} at lower
temperatures with the ``force free'' phenomenon {\it kept}. Within the
LD model, this $B$-dependence will be clarified only by taking
account of point disorder on the CuO planes inevitably contributing 
at lower temperatures. 
 
A vortex-glass transition temperature is determined as the 
temperature at which the uniform glass susceptibility $\chi_G({\bf
k}=0)$  
diverges on cooling, where $\chi_G({\bf k})$ is the Fourier
transformation of the correlation function\cite{FFH} 
$$G_G (md, \, \, {\bf R}_\perp) = d \sum_j \int {\rm d}y {\rm d}z 
{\overline {| \, 
< \psi^*_j({\bf r}_\perp) \, \psi_{j+m}({\bf r}_\perp + {\bf R}_\perp)> 
\, |^2}} \eqno(1.1)$$
defined consistently with the Lawrence-Doniach model 
$${\cal H}_{\rm LD} = d \sum_j \int {\rm d}y {\rm d}z \biggl[ 
\varepsilon_0
|\psi_j|^2 + \, \xi_0^2 \sum_{\mu=y,z} \biggl| \biggl( -{\rm i} 
\partial_\mu
+ {{2 \pi} \over {\phi_0}} A_{j, \mu} \biggr) \, \psi_j \biggr|^2 +
\Gamma^{-1} \biggl({{\xi_0} \over d} \biggr)^2 \, |\psi_j -
\psi_{j+1}|^2$$
$$ + \, {b \over 2} \, |\psi_j|^4 \biggr] \eqno(1.2)$$
$$=\int \! {\rm d}^3r \! \sum_{m = -\infty}^{+\infty} {\rm e}^{{\rm 
i}2
\pi m x/d} \biggl
[ \varepsilon_0 |\psi({\bf r})|^2 + \xi_0^2 \sum_{\mu = y,z} \biggl| 
\biggl(-{\rm i} \nabla_\mu + {{2 \pi} \over {\phi_0}} (B x {\hat y} +
\delta A_\mu({\bf r})) \, \biggr) \psi({\bf r}) \biggr|^2$$ 
$$+ \Gamma^{-1} \biggl({{\xi_0} \over d}
\biggr)^2 |\psi({\bf r}) - \psi({\bf r} + d {\hat x})|^2 
+ {b \over 2} |\psi|^4 \biggr], $$
where $d$ is the spacing between the superconducting layers lying in
$y$-$z$ plane, $\psi_j({\bf r}_\perp)$ with ${\bf r}_\perp = (y,  z)$
in the first representation of eq.(1.2) is the 
pair-field (superconducting order parameter) on the $j$-th
superconducting layer, $\varepsilon_0 \simeq (T - T_{c0})/T_{c0}$, 
$\Gamma$ the mass anisotropy, $\psi({\bf r})$ a spatially
continuous pair-field in the second representation of eq.(1.2), and 
$b \xi_0^{-2}$ is written, in terms of the zero temperature penetration 
depth $\lambda(0)$ or of the 2 D Ginzburg number (i.e., fluctuation
strength) $\varepsilon_G^{(2)}$, in the form $32 \pi^3
\lambda^2(0)/\phi_0^2 = 2 \pi d \varepsilon_G^{(2)}/k_{\rm B} T_{c0}$. 
Further, in the first representation we put ${\bf A}_j = B j d {\hat 
y} + \delta {\bf A}(jd, {\bf r}_\perp)$ by assuming ${\bf B} \parallel 
{\hat z} \perp {\bf c}$ ($= {\hat x}$), where $\delta {\bf A}$ is an
external gauge disturbance necessary 
in defining linear response quantities. In
obtaining the second representation, the Poisson summation formula was 
merely used with no approximation. Hence, the Josephson
vortices with no singular cores in the first representation are
expressed, for convenience, in the second representation as singular
vortices of $\psi({\bf r})$. In ref.5, a field-tuned sequence of
structural (mean field) transitions between various disorder-free 
vortex solids was examined in terms of the second representation. 

Point disorder on the superconducting
planes is introduced by adding, e.g., a random scalar potential 
term\cite{RI5} 
to eq.(1.2) 
$${\cal H}_{rp} = d \sum_j \int {\rm d}y {\rm d}z  \, u_j({\bf r}_\perp) \,
|\psi_j({\bf r}_\perp)|^2 \, , \eqno(1.3)$$ 
where the random potential $u_j$ has zero mean and Gaussian ensemble 
${\overline {u_j({\bf r}_{\perp 1}) 
\, u_l({\bf r}_{\perp 2})}} = \xi_0^2 \Delta \, \delta^{(2)}
({\bf r}_{\perp 1} - {\bf r}_{\perp 2}) \, \delta_{j, l}$. Here we did
not introduce other uncorrelated disorder terms because, as will be seen
later, our main results in this paper are independent of the presence of 
such additional terms. 

In $\S 2$, the model (1.2) with the term (1.3) is examined consistently
with ref.5 using the LLL approximation, and 
the glass susceptibility $\chi_G({\bf k})$ is examined to 
find a glass transition line of the Josephson vortex system, 
i.e., of ${\bf B} \perp {\bf c}$ case of the LD model by focusing on the 
nontrivial high field regime $B > \phi_0/(2 \pi d^2 \sqrt{\Gamma})$. 
In $\S 3$, it is demonstrated by examining tilt responses of the LD
model and of the closely related uniaxially periodic model\cite{Per} that
the resulting Josephson-vortex-glass (JG) phase should have a transverse
Meissner effect (TME) for a tilt perpendicular to the layers, while the 
response for a tilt within the layers (perpendicular to ${\bf c}$ axis)
remains nonsuperconducting. 
As intuitively expected by identifying the 
superconducting layer in the present case with a well-defined splayed 
plane\cite{RI2} formed by line-like defects, such an anisotropy of tilt 
responses is similar to that in the planar (Gaussian) splayed-glass 
phase examined in ref.4, although they are different in the origins of 
TME from each other. 
In $\S 4$, the resulting ${\bf B} \perp {\bf c}$ phase
diagrams to be realized in a manner depending on the material parameters 
are discussed in details, and consistencies with available
experimental data\cite{Kad,Ishiguro} are pointed out. 

\section{Glass Transition Line}

Let us begin with rewriting\cite{RI3} the LD model (1.2) (with no 
magnetic
screening) by assuming the continuous 
pair-field $\psi({\bf r})$ to be composed of
the LLL modes. The disorder terms, eq.(1.3), will be omitted for the 
moment. 
We will express $\psi$ as $\psi({\bf r}) = \sum_{Q, \, q_z}
\varphi(n, \, {\bf q}_\perp) u_Q(x,y) \exp({\rm i}q_z z)$ 
with the LLL eigenfunction
$u_Q(x,y) \propto \exp({\rm i}Q y - \sqrt{\Gamma}(x + Q r_B^2)/(2 r_B^2) 
\, )$ in the gauge ${\bf A} = B x {\hat y}$ and substitute it into the
second representation of eq.(1.2). Due to the layer structure with
period $d$, the momentum $Q$ is written\cite{RI3} as $Q = 
r_B^{-2} d n + q_y$ with $r_B = \sqrt{\phi_0/2 \pi B}$, where $|q_y| < 
d 
r_B^{-2}/2$, and the partial LLL degeneracy is measured by the integer
$n$ with the degree of degeneracy $N_d \equiv L_x/d$, where $L_\mu$
denotes the linear system size in $\mu$-direction. For the moment, 
a {\it constant} gauge disturbance 
$\delta {\bf A}$ will be assumed. Using the Poisson summation formula, 
the 
quadratic part of eq.(1.2) is expressed as 
$${\cal H}_2 = \xi_0^2 \sum_{Q, \, q_z} \biggl( \xi_0^{-2} 
\varepsilon_0
\, + \, \biggl< \biggl({{2\pi} \over {\phi_0}} \delta A_y - q_y + {d 
\over
{r_B^2}} m \biggr)^2 \biggr>_m $$
$$+ {2 \over {\Gamma d^2}} 
\biggl( 1 \, - \, {\rm e}^{-p/4}< \exp(- p(m + 1/2 - q_y r_B^2/d)^2) >_m 
\biggr) + \, \biggl({{2 \pi} \over {\phi_0}} \delta A_z - q_z \biggr)^2 
\,
\biggr) | \, \varphi(n, {\bf q}_\perp) \, |^2, \eqno(2.1)$$
where 
$$p = {{\sqrt{\Gamma} d^2} \over {r_B^2}} = {{2 \pi d^2 
\sqrt{\Gamma} \, 
B} \over {\phi_0}} \eqno(2.2)$$ 
is a dimensioless field playing important roles in ${\bf B} \perp {\bf
c}$, $\varphi(n, {\bf q}_\perp)$ was
rescaled in the way $\varphi /\sqrt{h(q_y)} \to 
\varphi$, $h^{-1}(q_y) = \sum_m \exp(-p(m - q_y r_B^2/d)^2)$, and
$<\!s(m)\!>_m$ denotes $ \, h(q_y) \times \sum_m [ \, s(m) \, \exp(-p (m 
- q_y
r_B^2/d)^2) \,]$. Assuming ${\rm e}^{-p/2} \ll 1$ 
and neglecting higher order terms
in $q_y$, eq.(2.1) is simplified in the form
$${\cal H}_2 \simeq \sum_{\bf Q} \biggl( \mu_0 + \xi_0^2 \biggl( q_z -
{{2 \pi} \over {\phi_0}} \delta A_z \biggr)^2$$ 
$$+ \xi_0^2 \biggl( q_y^2 (1 - 2 e^{-p/2}) - 2 q_y {{2 \pi} \over
{\phi_0}} \delta A_y (1 - 4 p e^{-p}) + \biggl({{2 \pi} \over {\phi_0}} 
\delta A_y \biggr)^2 \, \biggr) \, \biggr) \, | \, \varphi(n, {\bf 
q}_\perp)
\, |^2, \eqno(2.3)$$
where $\mu_0 \simeq \varepsilon_0 + 2 \Gamma^{-1} (\xi_0/d)^2 \, (1 - 2
{\rm e}^{-p/2})$. 
Hereafter, we assume the relation $\Gamma > 2 \xi_0^2/d^2$, 
promising\cite{RI3}
the presence of the $p > 1$ region in which the vortices tend to be 
confined between two neighboring layers. Since only the LLL modes 
were taken into account here, the gauge field 
$\delta A_y$ in $x$-$y$ plane perpendicular to ${\bf B}$, as in ${\bf B}
\parallel {\bf c}$ case, need not appear as the gauge-invariant 
combination $q_y - 2 \pi \delta A_y/\phi_0$ in ${\cal
H}_2$. However, in the present ${\bf B} \perp {\bf c}$ case of the LD
model, the $q_y^2|\varphi(n, {\bf q}_\perp)|^2$ 
and $q_y \delta A_y |\varphi(n,{\bf q}_\perp)|^2$ terms appear within
LLL even for a smaller $p$ as a consequence of the partial breaking of
the LLL degeneracy due to the layering. In particular, the latter term
implies that the linear responses to $\delta A_y$, at least at low
enough temperatures, can be defined within the LLL modes. 
Further, it is important
to note that the last three terms of eq.(2.3) are summarized in the 
form $\xi_0^2 (q_y - 2 \pi \delta A_y/\phi_0)^2$, just as the second 
term of (2.3), if the O(${\rm e}^{-p/2}$) 
corrections in eq.(2.2) are neglected by assuming a strong field or 
a large enough anisotropy. This recovery of the isotropy in $y$-$z$
plane in high $p$ should be expected in the LD model: Since, in this
model, the pair-field is defined only on the discrete superconducting 
layers, the vortices in $p > 1$ are confined between two neighboring 
layers, and consequently, the field-induced anisotropy on the layers
tends to be lost with increasing $p$. 
This recovery of the $y$-$z$ isotropy and the fact\cite{RI3} that 
the high field limit of the mean field transition line given by 
$\mu_0=0$ is the same as the exact result\cite{Klemm} suggest 
that the high field regime of the LD model may be well approximated 
by the present 
LLL approximation even in examining the response properties in $y$-$z$
plane. For these reasons, for later convenience, we will give the high 
field
approximation of eq.(1.2) including the $|\psi|^4$ term in the 
form\cite{RI3} 
$${\tilde {\cal H}_{\rm LD}} (\varphi) = \sum_{\bf Q} \biggl( \mu_0 + 
\xi_0^2 
\sum_{\mu=y, z} \biggl( q_\mu - {{2 \pi} \over {\phi_0}} \delta A_\mu
\biggr)^2 \biggr) \, |\varphi(n, {\bf q}_\perp)|^2 + \, {b \over {2 d
L_y L_z}}$$ 
$$ \times \sum_{{\bf Q}_1, {\bf Q}_2, {\bf Q}_3} \, V_0(n_1-n_3,
n_2 - n_3, \, q_{y,j}) \, \varphi^*(n_1, {\bf q}_{\perp,1}) \, 
\varphi^*(n_2, 
{\bf q}_{\perp,2})$$ 
$$\times \varphi(n_3, {\bf q}_{\perp,3}) \,
\varphi(n_1+n_2-n_3, {\bf q}_{\perp,1} + {\bf q}_{\perp, 2}-{\bf
q}_{\perp,3}), \eqno(2.4)$$
where 
$$V_0(n_1-n_3, n_2-n_3, \, q_{y, j}) = \sqrt{h(q_{y,1} + q_{y_2} -
q_{y_3}) \prod_{j=1,2,3} h(q_{y,j})} $$ 
$$\times {\overline v}(Q_1-Q_3) \,
{\overline v}(Q_2 - Q_3) \, \sum_m {\overline v}(Q_1 + Q_2 - 2dm/r_B^2)
\eqno(2.5)$$ 
is a bare vertex function with ${\overline v}(Q_i) 
= \exp(-p \, (n_i + q_{y,i} r_B^2/d)^2/2)$, and ${\bf Q}_i = Q_i {\hat
y} + q_{z,i} {\hat z} = (d n_i/r_B^2) {\hat y} + {\bf q}_{\perp, i}$. 

The above analysis is not modified by replacing the {\it constant} 
$\delta
{\bf A}$ assumed above with a $y$-dependent $\delta A_z$ and/or a 
$z$-dependent $\delta A_y$. This case is relevant to examining the tilt
response near and above the glass transition and will be discussed in $\S
3$. Rather, we will comment here on the superconducting part 
$\sigma_{s,\mu \mu}$ ($\mu=y$ or $z$) of dc (uniform) conductivities 
in the liquid region 
above $T_m$ where the pinning disorder term, eq.(1.3), may be neglected. 
According to the Kubo formula, it takes the form $\sigma_{s,\mu \mu}  
= - \partial \, \Upsilon_\mu({\bf q}=0, {\rm i}\Omega)/\partial
\Omega|_{\Omega \to +0}$, 
where $\Upsilon_\mu ({\rm i}\Omega)$ is a dynamical helicity modulus
defined by substituting a $\tau$-dependent gauge disturbance $\delta
A_\mu(\tau)$ into a quantum GL action\cite{RI1} 
consistent with eq.(1.2), $\tau$ 
the imaginary time, and $\Omega$ is an external Matsubara frequency. 
If the gauge field $\delta {\bf A}$ is spatially uniform, the 
derivation of a LLL quantum action with $\tau$-dependent $\delta {\bf
A}$ is the same as the above derivation of eq.(2.4). Then, since 
the gradient and gauge-dependent terms when $e^{-p} \ll 1$ are isotropic 
in $y$-$z$ plane, and the ${\bf q}_{\perp,l}$ dependence 
($l=1$, $2$, or $3$) of the bare vertex function $V_0$ of eq.(2.4) 
is negligible, it is
clear that the resulting $\sigma_{s,yy}$ and $\sigma_{s,zz}$ in such
high fields are the same as each other, implying the in-plane
``force-free'' behavior mentioned in $\S 1$. 

As is clear from the above 
derivation of eq.(2.4), this ``force-free'' behavior becomes more
accurate with increasing $p = 2 \pi B d^2 \sqrt{\Gamma}/\phi_0$, 
i.e., with increasing $B$ and/or $\Gamma$. 
This is consistent with the experimental fact\cite{Kwok,Iye2} that, in
the same field range, the in-plane ``force-free'' behavior 
is seen not in less anisotropic YBCO but only in strongly anisotropic 
BSCCO. In the liquid regime in $p \gg 1$, the in-plane conductivities
$\sigma_{s, \mu \mu}$ ($\mu=y$, $z$) are $B$-independent 
and well approximated by 
the fluctuation conductivity for 2 D and $B=0$ case, obtained by
neglecting the Kosterlitz-Thouless critical behavior, and have the form 
$$\sigma_{s, \mu \mu} \simeq {\pi \over {32 \mu_c R_q d}} \exp\biggl(-{2 
\over 
{\varepsilon_G^{(2)}}}\biggr) \, \exp\biggl({{T_{c0}} \over 
{\varepsilon_G^{(2)} \, T}} \biggl(1 - {{2 \xi_0^2} \over {d^2 \Gamma}}
\biggr) \biggr), \eqno(2.6)$$
where $R_q = \pi \hbar/2 e^2 = 6.45 ({\rm k}\Omega)$ is the 
resistance
quantum, $\mu_c$ is a constant of order unity, 
and a relation (2.21) (to be introduced later) was used here. 
Namely, the resistivity in the liquid regime in 
$p \gg 1$ yields the thermally activated behavior with {\it no} pinning
effect. 

In a previous paper\cite{RI4}, one of us has
ascribed the ``force free'' behavior to an extremely strong thermal 
fluctuation (equivalent to the vortex loop excitations) 
enhanced by the strong anisotropy of BSCCO, and has argued
the discrete layer structure not to be essential to this phenomenon. The 
feature suggested in Fig.2 of ref.11 that the $B$ dependence is lost 
with increasing $B$ indicates that the present explanation requiring the 
layer structure is more reasonable, and the assumption of an
extremely strong fluctuation, which cannot be described within the LLL 
approximation, seems to be unnecessary in understanding this resistive 
behavior.\cite{Kad}    

We return to detailing our theory. The isotropic
response in $y$-$z$ plane mentioned above, i.e., the ``force free'' 
behavior does not change by including any point disorder unless this
static disorder makes the supercurrent within the superconducting 
planes {\it anisotropically} random. Such a unusual possibility of 
disorder will not be considered here, and
we will focus on eq.(1.3) as a reasonable model of such an isotropic 
point disorder. After replicating the model (1.2) with 
the term (1.3), the replicated effective hamiltonian\cite{Lub} 
becomes ${\cal H}_{\rm eff} = \sum_{a=1}^n {\tilde {\cal
H}_{\rm LD}}(\varphi^{(a)}) 
+ \sum_{a, b=1}^n {\cal H}_{\rm ran}(\varphi^{(a)}, \varphi^{(b)})$, 
where 
$${\cal H}_{\rm ran}
(\varphi^{(a)}, \varphi^{(b)}) = {{- \Delta^{(2)}} \over 
{2 {\rm k}_{\rm B} T d L_y L_z}} \sum_{{\bf Q}_1, {\bf Q}_2, {\bf Q}_3} 
\,
V_0(n_1-n_3, n_2 - n_3, \, q_{y,j})$$ 
$$ \times \varphi^{(a) *}(n_1, {\bf q}_{\perp,1}) \,
\varphi^{(b) *}(n_2, {\bf q}_{\perp,2}) \, \varphi^{(a)}(n_3,
{\bf q}_{\perp,3}) \, \varphi^{(b)}(n_1+n_2-n_3, {\bf q}_{\perp,1} 
+ {\bf q}_{\perp,2} - {\bf q}_{\perp,3}). \eqno(2.7)$$
Using this replicated hamiltonian, the glass susceptibility within LLL 
consistent with ${\cal H}_{\rm eff}$ is written, after using the Poisson 
summation formula, in the form 
$$\chi_G({\bf k}) \simeq {\cal N} \sum_{{\bf Q}_1, {\bf Q}_2} e^{{\rm 
i}k_x
d(n_1 - n_2)} \, {\overline {<\varphi_{{\bf Q}_1} \, \varphi^*_{{\bf 
Q}_2}> 
\, <\varphi_{{\bf Q}_2 + {\bf k}_\perp} \, \varphi^*_{{\bf Q}_1 
+ {\bf k}_\perp}>}}, \eqno(2.8)$$
where ${\bf k}_\perp = k_y {\hat y} + k_z {\hat z}$, and ${\cal N}$ is 
a normalization factor weakly dependent on $k_y$. 
In obtaining eq.(2.8), we assumed again ${\rm e}^{-p} \ll 1$. 

Following the procedures used in ${\bf B} \parallel {\bf c}$
case\cite{RI1,RI6}, let us consider the perturbation series of eq.(2.8), 
in particular its irreducible part $I_{\rm ir}$. In the {\it limit} of 
weak 
disorder, $I_{\rm ir}$ consists of a single impurity line, 
carrying a strength $\Delta$, with vertex corrections determined only by 
the 
interaction (quartic) term of eq.(2.4). 
Essential features to $I_{\rm ir}$ can be seen in the 
study\cite{RI6,RI1} 
of $\chi_G(0)$ in the simplest 2 D limit of ${\bf B} 
\parallel {\bf c} \parallel {\hat z}$ 
which will be reviewed here. In this case, $I_{\rm ir}$ 
is a constant so that $\chi_G(0)$ is the simple geometrical series, 
$\chi_G(0) = 1 + I_{\rm ir} + I_{\rm ir}^2 + \cdot \cdot \cdot $. In a 
representation, corresponding to eq.(2.4), of hamiltonian in this 
${\bf B} \parallel {\bf c}$ case, $I_{\rm ir}$ obtained in ref.22 is 
expressed in the form 
$$I_{\rm ir}|_{{\rm 2d}, \perp} 
= {{x_R \, \Delta_{\rm eff}} \over {2 \pi N_v^{1/2}}}
\sum_{k_y, p, p'} v_{k_y} v_{p-p'} \, (\delta_{p, 0} - 2 x_R {\overline
V}(k_y|p) ) \, (\delta_{p', 0} - 2 x_R {\overline V}(k_y|p') ), 
\eqno(2.9)$$ 
where ${\overline V}(k_y|p) = N_v^{-1} \sum_{k_x} V_{\bf k} 
e^{-{\rm i}p k_x r_B^2}$, ${\bf k}$ is here a wave vector perpendicular
to ${\bf B}$, $V_{\bf k}$ is a fully
renormalized vertex part corresponding to the bare one $v_{k_x} 
v_{k_y}$ with $v_{k_\mu} = \exp(-k_\mu^2 r_B^2/2)$, $\Delta_{\rm eff} 
= \Delta/\varepsilon_G^{(2)}(T)$, with $\varepsilon_G^{(2)}(T) =
\varepsilon_G^{(2)} T/T_{c0}$, is 
the effective pinning strength\cite{RI1,RI6}, and $x_R$ is an 
interaction strength proportional to the squared pair-field
propagators\cite{RI1}. According to the mean field result and a more 
general
expression of the Abrikosov factor $\beta_A$, one notices\cite{RI1,RI6} 
that, in the limit of the perfect (square) vortex lattice,
$V_{{\bf k}_\perp}$ reduces to $(2 x_R)^{-1}( 1 - N_v \sum_{{\bf G} \neq =
0} 
\delta_{{\bf k}, {\bf G}} )$, i.e., 
$${\overline V}(k_y|p) \to (2 x_R)^{-1} ( \delta_{p,0} + \delta_{k_y, 0} 
- 
\sum_{G_2} \delta_{k_y, G_2} \sum_{G_1} e^{-{\rm i}p G_1 r_B^2} ), 
\eqno(2.10)$$ 
where ${\bf G} = (G_1, G_2)$ is a 
reciprocal lattice vector, and $N_v$ the total number of
vortices. Eq.(2.9) with eq.(2.10) simply becomes 
$I_{\rm ir}|_{{\rm 2d}, \perp} \propto N_v (\beta_A - 1)$. Since
the factor $\beta_A - 1$ is always positive, reflecting a spatially
varying $|\psi|$ in any vortex state, this prefactor $N_v$ in $I_{\rm
ir}$ indicates that, at the mean field level, a glass ordering 
occurs as soon as the vortices solidify on cooling. Namely, a spatial 
variation of $|\psi|$ (i.e., the fact $\beta_A > 1$) in ordered phases,
due to the presence of vortices, induces a glass ordering in nonzero
fields contrary to in zero field case (see $\S 1$). In a more
pinning-disordered case, the factor $N_v$ in the above relation is
replaced with the number of vortices in a correlation area, $N_{\rm
cor}$, to be realized below a (if any) solidification 
transition (or crossover) line. This interpretation was 
justified\cite{RI6} noting the fact that a term in $V_{{\bf k}_\perp}$
corresponding to the last term of eq.(2.10) is nothing but the 
structure factor measuring the vortex-positional ordering. 
In a dirty enough case, the above-mentioned term in $V_{{\bf k}_\perp}$, 
just like above the melting line, becomes a continuous 
function of $|{\bf k}_\perp|$ decaying with increasing $|{\bf k}_\perp|$ 
even below the pinning-free melting line, 
and the positional correlation length shrinks 
to a value of the order of $r_B$ (Note that, as far as the fluctuation
is weak or negligible, a nonzero positional correlation length is
well-defined). Then, the factor $N_{\rm cor}$ is found\cite{RI6} to 
decrease 
to a value of O(1), resulting in a decrease of the glass transition
point. It is this situation with $N_{\rm cor}$ of order unity which 
is accidentally described\cite{RI6} in the
simplest ladder approximation with {\it no}\cite{DF} vertex corrections 
to the 
impurity line. Further, a decrease of temperature, i.e., an increase of 
$B$, along the melting line leads to an increase of $\Delta_{\rm
eff}$ and hence, corresponds to an enhancement of {\it microscopic} 
pinning. 
The $B$-$T$ phase diagram thus obtained\cite{RI1} in ${\bf B} \parallel
{\bf c}$ is consistent with recent data\cite{Koban} 
of twin-free YBCO (see $\S 4$). 
This example for ${\bf B} \parallel {\bf c}$ case 
indicates that an inclusion of a vertex
correction, implying the vortex positional correlation, to the
impurity line organizing $\chi_G$ 
is indispensable to understanding a field dependence of
the glass transition temperature for a fixed pinning strength. 
 
Now, to perform the corresponding analysis for the present ${\bf B} 
\perp {\bf c}$ case, let us first examine the fully-renormalized vertex 
part in the ideal limit of the perfect solid in $p > 1$ which was called 
the $w=1$ regular solid in Ref.5. The mean field 
result of $\beta_A$ of this solid takes the form\cite{RI3} 
$$\beta_A^{(1,1)} 
= \sum_{m, n} (-1)^{m \, n} \, V_0(m, n, \, q_{y,j}=0). 
\eqno(2.11)$$ 
In the cases including thermal fluctuations, $\beta_A$ is 
expressed as $f(0) + 1$, where $f({\bf R})$ is the
density-density correlation function\cite{RI9}, and takes the
form\cite{RI3} in ${\bf B} \perp {\bf c}$ 
$$f({\bf R}) = N_d^{-1} ( \, \sum_n e^{-p n^2} \sum_{{\bf q}_\perp} 
G({\bf q}_\perp) \, )^{-2} \sum_{{\bf Q}_1, {\bf Q}_2, {\bf Q}_3} \,
V_0(n_1 - n_3, n_2 - n_3 - M, \, q_{y,j}=0)$$ 
$$ \times ( \, < \, \varphi^*(n_1, {\bf 
q}_{\perp,1}) \, \varphi^*(n_2, {\bf q}_{\perp,2}) \, \varphi(n_3, 
{\bf q}_{\perp,3}) \, \varphi(n_1+n_2-n_3, {\bf q}_{\perp,1} 
+ {\bf q}_{\perp,2} - {\bf q}_{\perp,3}) \, >$$
$$ - < |\varphi(n_3, {\bf
q}_{\perp, 3})|^2 > < |\varphi(n_2, {\bf q}_{\perp, 2})|^2 >
\delta_{Q_1, Q_3} \, ) \,\, e^{{\rm i}(Q_3 - Q_1) Y}, \eqno(2.12)$$
where ${\bf R} = (M d, Y)$ with an integer $M$. 
Noting that, due to the partial LLL degeneracy\cite{RI3}, the 
fully-renormalized interaction term takes the form 
$${\cal H}_4^{\rm ren} = {b \over {2 d L_y L_z}} \sum_{{\bf Q}_1, {\bf 
Q}_2, 
{\bf Q}_3} V(n_1-n_3, n_2-n_3, \, {\bf q}_{\perp,j}) \, \varphi^*(n_1, 
{\bf 
q}_{\perp,1}) \, \varphi^*(n_2, {\bf q}_{\perp,2}) \, \varphi(n_3, {\bf
q}_{\perp,3})$$
$$ \times \varphi(n_1+n_2-n_3, {\bf q}_{\perp,1}+{\bf
q}_{\perp,2}-{\bf q}_{\perp,3}), \eqno(2.13)$$
and writing\cite{RI3} eq.(2.12) in terms of this renormalized 
vertex $V(m,n, \, {\bf q}_{\perp,j})$ and comparing it with eq.(2.11), 
one finds that eq.(2.11) is recovered from eq.(2.12) when 
$${{2b} \over {d L_y L_z}} \sum_{{\bf q}_\perp} G({\bf q}_\perp) 
G({\bf q}_\perp + {\bf k}_\perp) \, V(n_1, n_2, \, {\bf q}_\perp, 
{\bf q}'_\perp, {\bf k}_\perp) \simeq (-1)^{1+n_1 n_2} + \delta_{n_1, 0} 
+
\delta_{n_2, 0}, \eqno(2.14)$$
in the limit of vanishing ${\bf q}'_\perp$ and ${\bf k}_\perp$. Notice
that eq.(2.14) corresponds to eq.(2.10) in ${\bf B} \parallel {\bf c}$. 
By noting that the renormalized version of 
eq.(2.7) simply becomes 
$$ {\cal H}_{\rm ran}^{\rm ren}(\varphi^{(a)}, \varphi^{(b)}) 
= {{- \Delta^{(2)}} 
\over {2 {\rm k}_B T d L_y L_z}} 
\sum_{{\bf Q}_1, {\bf Q}_2, {\bf Q}_3} 
\, \sum_{n_4, n_5} V_0(n_1-n_3, n_4 - n_5, \, q_{y,j})$$
$$ \times C_v(n_3-n_5, n_1-n_3,
\, {\bf q}_{\perp,1}, {\bf q}_{\perp,3}) \, C_v(n_2-n_4, n_1-n_3, 
\, {\bf q}_{\perp,2}, 
\, {\bf q}_{\perp,1}+{\bf q}_{\perp,2}-{\bf q}_{\perp,3})$$ 
$$ \times \varphi^{(a) *}(n_1, {\bf q}_{\perp,1}) \,
\varphi^{(b) *}(n_2, {\bf q}_{\perp,2}) \, \varphi^{(a)}(n_3, {\bf q}_
{\perp,3}) \, \varphi^{(b)}(n_1+n_2-n_3, {\bf q}_{\perp,1} 
+ {\bf q}_{\perp,2} - {\bf q}_{\perp,3}), 
\eqno(2.15)$$
where 
$$C_v(n, m, \, {\bf q}_{\perp,1}, {\bf q}_{\perp,2}) 
= \delta_{n,0} - {{2b } \over {d L_y
L_z}} \sum_{{\bf q}_\perp} G({\bf q}_\perp + {\bf q}_{\perp,1}) G({\bf
q}_\perp + {\bf q}_{\perp,2}) \, V(n, m, \, {\bf q}_\perp, {\bf
q}_{\perp,1}, {\bf q}_{\perp,2}), \eqno(2.16)$$ 
and applying eq.(2.14) to this, the expression 
of eq.(2.15) simply becomes in clean {\it limit} 
$${\cal H}_{\rm ran}^{\rm ren}(\varphi_\alpha, \varphi_\beta) 
= {{- \Delta^{(2)}} \over {2 {\rm k}_{\rm B} T d L_y L_z}} \, N_d 
\sum^{(n_1 \neq n_3)}_{{\bf Q}_1, {\bf Q}_2, {\bf Q}_3} 
(-1)^{(n_1-n_3)(n_2-n_3)} \sum_n
V_0(n_1-n_3, n, \, q_{y,j}=0)$$ 
$$\times (-1)^{n(n_1-n_3)} \, \varphi^*_\alpha(n_1, {\bf q}_{\perp,1}) \,
\varphi^*_\beta(n_2, {\bf q}_{\perp,2}) \, \varphi_\alpha(n_3, {\bf q}_
{\perp,3}) \, \varphi_\beta(n_1+n_2-n_3, {\bf q}_{\perp,1} 
+ {\bf q}_{\perp,2} - {\bf q}_{\perp,3}). \eqno(2.17)
$$
The condition $n_1 \neq n_3$, indicated above, in the summations of 
eq.(2.17)
corresponds to a fact that the irreducible vertex $I_{\rm ir}$ of 
$\chi_G$
is proportional to ${\rm e}^{-p}$ ($\propto \beta_A - 1$) 
(see the expression of
$r_G$ in eq.(2.19)) and implies that, as in ${\bf B} \parallel {\bf c}$, 
spatial variations of $|\psi|$ and of the supercurrent in a direction 
perpendicular to ${\bf B}$, which is the $y$-direction parallel to the 
superconducting planes in ${\bf B} \perp {\bf c}$, in the expected
solid state are indispensable to a vortex glass ordering. 
The prefactor $N_d = L_x/d$ in eq.(2.17) corresponds to the prefactor 
$N_v$ 
in the resulting $I|_{{\rm 2d}, \perp}$ ( $\simeq N_v (\beta_A - 1)$ ) 
in ${\bf B} \parallel {\bf c}$ case and appeared here owing to the
ideal assumption such that the vertex correction to the impurity line
can be expressed with {\it no} pinning
disorder. Due to this large factor, the glass transition in clean 
limit of ${\bf B} \perp {\bf c}$ case also occurs, as 
in ${\bf B} \parallel {\bf c}$ case, as soon as the vortices solidify 
at the disorder-free melting line $T_m(B)$. 

Physically, the $N_d$-factor of eq.(2.17) 
implies the number in $x$-direction of 
Bragg spots, appearing below the melting line, at nonzero $k_y$'s 
(As indicated in eq.(4.5) of ref.5, the ``Bragg peaks'' at zero $k_y$ 
already appear above the melting line, merely reflecting the original 
layer structure, and are unrelated to an ordering in the
vortex state). To justify this identification, one has only to notice 
that the origin of this $N_d$-factor is a trivial (or, unlimited)
summation of $n_4$ or $n_5$ in eq.(2.15) 
and that, in eq.(2.12), $(n_2 - n_3) d$ plays, just like $n_4 d$ and
$n_5 d$ in eq.(2.15), a role of a coordinate perpendicular to the 
layers, while $Q_1 - Q_3$ is a wave vector in $y$-direction
parallel to the layers there. 
Since a positional correlation length just 
below $T_m$ in {\it real} systems shrinks with effectively increasing 
disorder, the $N_d$-factor arising by assuming the limit of weak 
pinning should be replaced, in more realistic cases, with a 
$\Delta$-dependent 
smaller factor. Diagrammatically, this implies that, for such a highly 
disordered case, the impurity lines within the vertex correction 
to each impurity line organizing the ladder of $\chi_G$ 
are not negligible any longer. Although it is not easy to find a $\Delta$
dependence of the vertex correction within the framework of the 
present work, we expect by following the corresponding analysis in 
${\bf B} \parallel {\bf c}$ case that, in such a disordered system, the 
first term of r.h.s. of eq.(2.14) will become a continuous function 
of $|n_1|$ and $|n_2|$ decaying with increasing $|n_1|$ or $|n_2|$
(This can be recognized through a perturbative computation valid above 
$T_m$). Then, the $n_4$ and $n_5$ 
summations in eq.(2.15) becomes convergent, and a prefactor in 
the corresponding one to eq.(2.17) does not depend on a system-size 
any longer. Hereafter, we will replace $N_d$ in eq.(2.17) with a smaller
number $N_c(\Delta)$ by imagining a moderately disordered 
system. Similarly to that in ${\bf B} \parallel {\bf c}$ case, this
$N_c$ physically corresponds to a (dimensioless) correlation length 
perpendicular to the layers. In such a moderately disordered system, 
the first order solidification transition at $T_m$ is weakened with 
increasing $p$ and {\it without} decreasing $T$ because the rigidity 
essential to the solidification, i.e., the shear rigidity, is 
exponentially 
small,\cite{RI3} and hence, the vortex system in higher $p$ is likely to 
be more susceptible\cite{com1} to the pinning disorder. Namely, besides 
that in ${\bf B} \parallel {\bf c}$ (see the preceding paragraph), 
this mechanism also contributes to disordering the solid due to an 
increase of $B$. Hence, the first order solidification line may 
terminate at a $p$-value $p_c$ before entering the $p \gg 1$ region. 
In a dirtier system with $p_c < 1$, the layering effect in $p < p_c$ is 
a 
small correction\cite{RI3} to the solidification, 
and thus, phenomena near $T_m$ in such $p$'s will be similar 
to those in ${\bf B} \parallel {\bf c}$. 

It is straightforward to derive an expression of $\chi_G({\bf k})$ in 
terms of eq.(2.17), 
which takes the form $\chi_G({\bf k}) = \int_{{\bf q}_\perp}
G_{{\bf q}_\perp} G_{{\bf q}_\perp + {\bf k}_\perp} \, (1 -
I_{\rm ir}({\bf k}) \, )^{-1}$ with 
$$I_{\rm ir}({\bf k}) = {\Delta \over d} N_c \int_{{\bf q}_\perp} 
G_{{\bf
q}_\perp} \, G_{{\bf q}_\perp + {\bf k}_\perp} \, \sum_{n \neq 0} \sum_m 
(-1)^{mn} \, V^{(0)}(m, n, \, q_{y,j}) \, {\rm e}^{{\rm i}dn k_x}. 
\eqno(2.18)$$ 
Here $G_{{\bf q}_\perp} = (\, \mu + \xi_0^2 \, {\bf q}_\perp^2 \, 
)^{-1}$
is the propagator in $p \gg 1$ 
of the pair-field fluctuation in LLL. Keeping the
lowest order terms in ${\rm e}^{-p}$ in the $n$, $m$-summations of 
eq.(2.18), 
the glass susceptibility is given by 
$$\chi_G({\bf k}) \propto \biggl( \, r_G + {{\xi_0^2} \over {6 \mu}}{\bf
k}_\perp^2 + 1 - {\rm cos} \biggl(2 k_x d \biggr) \, \biggr)^{-1}, 
\eqno(2.19)$$
where $r_G = \alpha - 1$ with $\alpha^{-1} = \Delta N_c \exp(-2p)/(2 
\pi 
\mu) $. If we focus on the $p$-region in which, due to point disorder, a 
first order transition\cite{RI3} accompanied by both the positional and 
(ordinary) superconducting orderings does not occur at $T_m \simeq 
T_{sc}$
(see below) any longer, 
the renormalized mass $\mu$ must remain positive (noncritical) even 
in $T < T_m$ because, as reflected in the expression (2.4), 
the LLL superconducting fluctuation is two-dimensional in nature. 
Owing to eq.(2.14), the mass-renormalization can be 
performed in the same manner as in ref.22, and one easily finds 
$$\mu = \mu_0 + 2 \pi \varepsilon_G^{(2)}(T) \, (\sum_m {\rm e}^{-p 
m^2})^2  
\int_{{\bf q}_\perp} {1 
\over {\mu + {\bf q}_\perp^2}} \biggl(\beta_A - \Delta_{\rm eff} \, 
(\beta_A -1) \biggr).  \eqno(2.20)$$
Since $\beta_A - 1 \sim $ O(${\rm e}^{-p}$) $\ll 1$ due to the
vortex-confinement between neighboring layers, we can neglect the 
last term of eq.(2.20) even at low enough temperatures so that 
the solution becomes $\Delta$-independent. The resulting $\mu$ can 
be expressed below 
the {\it mean field} phase boundary in the form 
$$\mu \simeq {{\varepsilon_G^{(2)}(T)} \over 2} \exp\biggl({{2 {\tilde 
\mu}_0} \over {\varepsilon_G^{(2)}(T)}} \biggr). \eqno(2.21)$$ 
Here ${\tilde \mu}_0 = \mu_0 + \varepsilon_G^{(2)}(T) \, {\rm ln}(2
\mu_c/\varepsilon_G^{(2)}(T))/2$, and $\mu_c$ is a cutoff constant of 
O($1$). 

Using this $\mu$-expression, the uniform glass susceptibility 
$\chi_G({\bf k}=0)$ is found to diverge 
at the Josephson-vortex-glass (JG) transition temperature 
$T_{JG}(B)$, where 
$${{T_{c2} - T_{JG}} \over {T_{JG}}} = 2 \pi \, \varepsilon_G^{(2)} \,
{{d^2 \sqrt{\Gamma}} \over {\phi_0}} ( \, B - \, B_c^{({\rm 
in})}(\Delta) \, ) 
, \eqno(2.22)$$
where $\varepsilon_G^{(2)} = 16 \pi^2 \lambda^2(0) k_{\rm B} 
T_{c0}/(\phi_0^2 d)$ is the 2 D fluctuation strength defined in $\S 1$,
and $T_{c2}$ is the mean field transition line. For simplicity, 
the correlation volume $N_c$ included in the characterstic field 
$$B_c^{({\rm in})}(\Delta) \simeq {{\phi_0} \over {4 \pi d^2 
\sqrt{\Gamma}}}
{\rm ln}\biggl({{ \, \Delta N_c(\Delta_{\rm eff})}
\over {2 \pi \mu_c}} \biggr),  \eqno(2.23)$$
which is the only measure of a disorder strength in eq.(2.22), is 
assumed here not to be $B$-dependent (see, however, $\S 4$). In very
clean systems, $B_c^{({\rm in})}$ becomes quite large, 
while it decreases with increasing disorder and may become negative in
moderately disordered case because the inequality,
$\varepsilon_G^{(2)}$, $\Delta < 1$, is expected even in real systems 
with strong thermal fluctuation such as BSCCO. As clearly seen in
the above discussion, eq.(2.22) is well defined in $B > B_c^{({\rm 
in})}$ and
in $T_{JG} < T_m$. The JG transition temperature $T_{JG}$ approaches 
the melting line $T_m$ as $B$ decreases and 
approaches $B_c^{({\rm in})}$, or equivalently, 
as the disorder becomes weaker
at a fixed $B$ value, i.e., when $N_c$ grows and approaches $N_d$.  
Oppositely, an increase of $B$ leads to a decrease of $N_c$, and hence, 
$T_{JG}(B)$ goes away from $T_m$. In high enough fields, $N_c(\Delta)$
reduces to a constant of order unity as in ${\bf B} \parallel {\bf c}$, 
and eq.(2.22) is simplified in the form
$${{{\tilde T}_{c2} - T_{JG}(B)} \over {T_{JG}(B)}} =
{{\varepsilon_G^{(2)}} \over {1 - \varepsilon_G^{(2)} {\rm
ln}\sqrt{\Delta/2 \pi}}} \, {{2 \pi d^2 \, \sqrt{\Gamma}} \over
{\phi_0}} \, B, \eqno(2.24)$$
where ${\tilde T}_{c2} = T_{c0}/(1 - \varepsilon_G^{(2)} {\rm ln}
\sqrt{\Delta/2 \pi})$. 
Note the weak dependence of $T_{JG}$ on the pinning strength $\Delta$. 

It is valuable to point out that, except a difference in the numerical
prefactor of order unity multiplying $B$, eq.(2.22) is of the same form
as a $B$-$T$ line following from the {\it disorder-free} Lindemann 
criterion 
$$< s_y^2 >_{\rm har} \sim {{r_B^4} \over {d^2}}, \eqno(2.25)$$
if $B_c^{({\rm in})}$ in eq.(2.21) is replaced 
by $B_{\rm el}^{({\rm in})} \simeq - (\phi_0/(2 \pi d^2 
\sqrt{\Gamma}) \, ) {\rm ln} [{\rm e}^4/(|\mu_0| \pi^4)]$. Here, $s_y 
= r_B^2 \partial_x \chi$ (see ref.26) is the displacement fluctuation 
parallel to the layers, and $\chi$ is the harmonic phase fluctuation
around the $w=1$ regular solid, i.e., the 
disorder-free ground state in $p > 1$, with its fluctuation energy 
$$\delta {\cal H}_{LD}^{(w=1)} = {{k_{\rm B} T_{c0} |{\tilde \mu}_0|}
\over {2 \pi \varepsilon_G^{(2)} d}} \int {\rm d}^3r \biggl( \, 
\sum_{\mu=y,z} 
(\partial_\mu \chi)^2 + \xi_0^{-2} |{\tilde \mu}_0| \, {\rm e}^{-p} \, 
(d^2 
\partial_x^2 \chi \, )^2 \, \biggr) \eqno(2.26)$$ 
(see eq.(3.8) of ref.5). The thermal average $<......>_{\rm har}$ in
eq.(2.25) is defined in terms of this fluctuation energy. Further, we
used in eq.(2.25) the fact that the average spacing between the vortices 
in $y$ direction of the $w=1$ solid is $2 \pi r_B^2/d$. 

To simplify our discussion on the result (2.22), we assume for a moment
$B_c^{({\rm in})} \ll \phi_0/(2 \pi d^2 \sqrt{\Gamma} 
\, \varepsilon_G^{(2)})$
(i.e., $\varepsilon_G^{(2)} {\rm ln}\sqrt{\Delta/2 \pi} \ll 1$) and 
hence, identify eq.(2.22) with 
$$T_{c2} - T_{\rm el}(p) \simeq T_{\rm el}(p) \, \varepsilon_G^{(2)}
\, p \, .  \eqno(2.27)$$ 
This $T_{\rm el}(B)$ decreases 
and approaches zero with increasing $p$. On the other hand,
the superconducting transition in each layer in the disorder-free case
may occur at a nozero temperature even in large enough $p$ ($\propto 
\sqrt{\Gamma}$) because the 2 D Meissner response is 
permitted\cite{KTB}. Thus, the disorder-free superconducting transition 
point $T_{\rm sc}$ should exist far above $T_{\rm el}$, 
at least in large enough $p$, 
and the only possible parameter-dependence determining such a
$T_{\rm sc}$  will be 
$$T_{c2} - T_{\rm sc} \simeq T_{\rm sc} \,
\varepsilon_G^{(2)}. \eqno(2.28)$$
Further, the solidification (melting)
transition at $T_m$ was argued in ref.5, as in ${\bf B} \parallel
{\bf c}$ case, to be of first order and simultaneously the
superconducting transition, implying that $T_m = T_{\rm sc}$, 
as far as the
thermally excited vortex loops are negligible (see Note added in poof in
ref.5). Thus, if the first order freezing transition line at $T_m$ 
terminates, 
for instance, due to the point disorder, somewhere in the $B$-$T$
phase diagram of a sample, the {\it ordinary} superconducting 
ordering in ${\bf B} \perp {\bf c}$ will not occur 
at any nonzero temperature in this sample.  

However, a possibility (in the disorder-free case) 
of a splitting between the superconducting and
melting transitions (i.e., $T_m < T_{\rm sc}$) was not excluded there, 
at least if going beyond the LLL approximation. 
If $T_m < T_{\rm sc}$, this melting transition, occuring between two 
superconducting phases, is associated only\cite{RI3} 
with the positional ordering 
across the layers, and $T_m$ will decrease with increasing $p$. A
reasonable guess of such a $T_m$ will be that $T_m \propto T_{\rm 
el}(p)$. 
However, it was argued\cite{KL} that a $B$-$T$ line obtained in terms of 
a Lindemann criterion neglecting vortex displacements across the layers 
should be far below a true melting transition in ${\bf B}\perp{\bf c}$. 
Actually, such a possibility that $T_m \simeq
T_{\rm el}(p) < T_{\rm sc}$ in $p > 1$ is unreasonable 
even from another point of view. To explain this, let us first point out 
that the expression (2.27) on $T_{\rm el}$ is {\it independent} of the
starting model and valid in the phase-only approximation, 
as far as the interlayer shear modulus is small 
like\cite{IK} $\sim {\rm e}^{-p}$, 
because the only model-dependence appears, in such a harmonic
fluctuation energy, as a prefactor of the interlayer shear energy, while 
such a prefactor is reflected merely logarithmically in a quantity 
corresponding to $B_{\rm el}^{({\rm in})}$ or $B_c^{({\rm in})}$ and 
hence, 
is negligible for large enough $p$. On the other hand, 
the melting line $T_m^{(c)}(B)$ in $p \ll 1$, 
where the layering effect may be negligible, will be well approximated 
by the
corresponding 
curve of the anisotropic GL model\cite{RI7} in ${\bf B} \perp {\bf c}$, 
which is, in the phase-only approximation, given by 
$$T_{c2} - T_m^{(c)}(p) \simeq T_m^{(c)}(p) \, \varepsilon_G^{(2)} 
\, \sqrt{p}, \eqno(2.29)$$ 
except a numerical prefactor of order unity of r.h.s. (see eq.(2.18) 
of ref.26). Thus, as shown in Fig.1, just three curves expressed by 
eqs.(2.27) to (2.29) are expected in the phase-only approximation of the 
LD model. Note that the point $p \sim 1$, at which the three curves
merge with each other, is a kind of dimensional (3 D to 2 D) crossover 
of the melting line in ${\bf B} \perp {\bf c}$ 
and corresponds
to the field\cite{RI8} $B'_{dc} \sim \phi_0/d^2 \Gamma$ in ${\bf B} 
\parallel {\bf c}$ case. Such a crossover to the 2 D regime 
in ${\bf B} \parallel {\bf c}$ 
is expected to occur with increasing $\Gamma$ in order to avoid a
unlimited enhancement of thermal fluctuation (Note that the $B=0$
fluctuation strength $\sqrt{\varepsilon_G^{(3)}} \propto
\varepsilon_G^{(2)} \sqrt{\Gamma}$ in 3 D grows unlimitedly as $\Gamma$ 
increases). Besides this, in the present ${\bf B} 
\perp {\bf c}$ case, we have not only such a {\it quantitative}
field-induced enhancemnet of fluctuation but also a {\it qualitative} 
field-induced reduction of the fluctuation due to the (1 D to 2 D) 
rise\cite{RI3} of its dimensionality. Thus, it is quite
unreasonable to expect that $T_m$ in ${\bf B} \perp {\bf c}$ and $p \gg 
1$
would be given by $T_{\rm el}(p)$ far below (the extrapolated)
$T_m^{(c)}$. 

Now, it will not be difficult any longer to understand why the 
vortex-glass 
transition position $T_{JG}(p)$ in $p \gg 1$ is related to 
the {\it pinning-free} Lindemann 
criterion (2.25). First, the small factor $\sim {\rm e}^{-p}$ is common 
both
to the interlayer shear modulus in the elastic energy 
and the prefactor in $I_{\rm ir}$ of the glass susceptibility (see
eq.(2.18)), and its origin is a weak spatial variation of $|\psi|$ 
in $y$-direction, i.e., a (small but) positive $\beta_A - 1$, 
in the solid state. The relation $\beta_A - 1 
\sim {\rm e}^{-p}$ may remain unchanged as far as the disorder 
is weak so that a granular structure, i.e., a spatial variation of 
$|\psi|$ {\it induced by} $\Delta$ at low enough $T$, on the layers is 
invisible. Further, the rise, due to the layering, of the dimensionality
of superconducting fluctuation is also reflected in the dispersion of
the elastic energy, and the resulting exponential $T$-dependence,
combining with the above-mentioned exponential $p$-dependence, makes a
role of disorder in the resulting transition line weak logarithmically. 
The only effect of the point disorder on the vortex elasticity near
$T_m$ and above $T_{JG}$ (i.e., in a vortex-slush regime\cite{RI1,Wor}) 
is regarded as a renormalization on the
interlayer shear modulus $\propto {\rm e}^{-p}$. When combined with the
model-independence of the $T_{\rm el}(p)$-line, this 
correspondence between $T_{JG}(p)$ and 
the elastic temperature $T_{\rm el}(p)$ convincingly 
suggests that the behavior $T_{c2} - T_{JG}(B) \sim T_{JG}(B) \,
\varepsilon_G^{(2)} \, p$ may be generally valid irrespective of our 
use of the LLL approximation. 

\section{Responses Quantities near $T_{JG}$} 

In this section, we examine linear response properties near and above
the transition point $T_{JG}$. As mentioned in relation to eq.(2.4), 
the physical quantities in the vortex liquid region with ${\rm e}^{-p} 
\ll 1$ 
are insensitive to the field $B$, reflecting the confinement of vortices 
in the spacings between two neighboring layers in $p \to \infty$, 
while in the vortex-slush region\cite{RI1,Wor} above $T_{JG}$ but below
$T_m$ they should depend on $B$ due to the remarkable $B$-dependence 
of $T_{JG}(B)$ given by eq.(2.22). 

This picture on the $B$-dependence also applies to the 
in-plane resistivities $\rho_{\mu \mu}$ 
($\mu=y$ or $z$) $= 1/(\sigma_n + \sigma_{s,\mu \mu})$, where
$\sigma_n$ is the normal part (quasiparticle contribution) of the
in-plane conductivity and reasonably assumed to be $\mu$-independent
(i.e., isotropic in $y$-$z$ plane). Since, just as in the case with no 
disorder (see the discussion leading to eq.(2.6)), the gauge-coupling 
and the gradient terms in ${\rm e}^{-p} \ll 1$ is approximated in an
isotropic form in $y$-$z$ plane, and the ${\bf q}_\perp$ dependence of 
the {\it bare} four-point vertices in the replicated hamiltonian in such 
high fields will merely play secondary roles, the total conductivities 
in 
high fields ${\bf B} \perp {\bf c}$ are also isotropic 
in $y$-$z$ plane. Namely, we have the ``force free'' behavior 
in ${\rm e}^{-p} \ll 1$ even when the point disorder is not
negligible. Further, the recovery of the gauge-coupling form in $y$-$z$
plane (see eq.(2.4)) within LLL also implies that this gauge-invariant
gradient operates on the glass order parameter formed within LLL. This
enables\cite{RI2} us to, just like the conductivity\cite{RI2} 
parallel to ${\bf B}$ 
in ${\bf B} \parallel {\bf c}$ case, anticipate the 
diverging behaviors of $\sigma_{s,\mu \mu}$ near $T_{JG}$ according to 
the 
scaling argument same as in the zero field transition, and consequently, 
we expect the critical behavior 
$$\sigma_{s,\mu \mu} \sim (T-T_{JG}(B))^{\nu_J(1 - z_J)}, \eqno(3.1)$$ 
where $z_J (> 4)$ and $\nu_J (> 1/2)$ are, respectively, the dynamical
exponent and the exponent of the correlation length, $\xi_J(T) \sim (T - 
T_{JG}(B))^{-\nu_J}$, describing the JG critical
behavior. In $p \gg 1$, the behavior (2.6) far above $T_{JG}$ (but 
below $T_{c0}$) is expected to, on cooling, smoothly change close to
$T_{JG}$ into the behavior (3.1) which is also isotropic in $y$-$z$
plane. Note, however, that, due to the $B$-dependence of $T_{JG}(B)$, the
in-plane resistivity should show a $B$-{\it dependent} ``force free''
behavior near $T_{JG}$. 

We will focus in the remainder of this section on 
the static responses to a tilt of the 
applied field which are measures of the presence or absence of a
transverse Meissner effect (TME) in the resulting glass (JG) phase. 
This static tilt response is a key quantity for distinguishing\cite{RI2} 
between various glass phases 
and is in this sense more nontrivial than the resistivities which vanish 
at a glass transition irrespective of the current direction. The static 
tilt response $\Delta C_{44,x}$, corresponding to the single vortex part 
of a tilt modulus, in the $x$-direction is defined\cite{RI2,RI8} by 
$$\Delta C_{44,x} = B^2 {{\Upsilon_y (k_x = k_y = 0, \, k_z)} 
\over 
{k_z^2}}\biggl|_{k_z=0}, \eqno(3.2)$$
where $\Upsilon_y(k_x=k_y=0, \, k_z)$ is the helicity modulus 
defined by 
a gauge disturbance $\delta A_y(z)$ dependent only on $z$, and the
corresponding $\Delta C_{44,y}$ is similarly defined in terms of $\delta 
A_x(z)$. Another kind of 
tilt moduli are defined in terms of $\delta A_z$ in the 
form\cite{RI8}
$$C'_{44,x} = B^2 {{\Upsilon_z(k_x=0, k_y, k_z=0)} \over 
{k_y^2}}\biggl|_{k_y=0}, \eqno(3.3)$$
which is a response to $\delta A_z(y)$. In the disorder-free 
liquid regime in ${\bf B} \parallel {\bf c}$, $\Delta C_{44, \mu}$ 
coincides\cite{RI8} with $C'_{44, \mu}$, and $\Delta C_{44, x}$ in the
(disorder-free) liquid regime in $p \gg 1$ is expected to, just as
$\sigma_{s, \mu \mu}$ (see eq.(2.6)), grow exponentially on 
cooling, i.e., 
$$\Delta C_{44,x} \sim \exp(T_{c0} (\varepsilon_G^{(2)})^{-1}/T). 
\eqno(3.4)$$ 

First, we will anticipate a critical behavior of $\Delta C_{44,x}$ with
no detailed calculation and simply by comparing with $C'_{44,x}$. To do
this, one only has to note that the derivation of the quadratic terms of 
eq.(2.4), in which $\delta A_\mu$ ($\mu=y$ and $z$) are constant, is
trivially extended to the case with a $z$-dependent $\delta A_y$ 
and a $y$-dependent $\delta A_z \sim \delta a(q_y) {\rm e}^{{\rm i}q_y 
y}$ with $|q_y| \ll d/r_B^2$. Namely, in this case the quadratic term of 
eq.(2.4) is replaced with 
$${\tilde {\cal H}}_{{\rm LD}, \, 2}(\varphi) = \int {\rm d}y {\rm d}z 
\sum_n \biggl( 
\xi_0^2 \biggl| \biggl(- {\rm i}  
\partial_y - {{2 \pi} \over {\phi_0}} \delta A_y(z) \biggr) \varphi(n,
{\bf r}_\perp) \biggr|^2 + \mu_0 |\varphi(n, {\bf r}_\perp)|^2 $$
$$ + \xi_0^2 \biggl| \biggl(- {\rm i} \partial_z 
- {{2 \pi} \over {\phi_0}} \delta A_z(y) \biggr) \varphi(n, {\bf 
r}_\perp) \biggr|^2 \biggr), \eqno(3.5)$$
where the real space representation for $\varphi$-fields is 
defined here by assuming any $y$-dependence to have wavelengths longer 
than $r_B^2/d$. 
As explained elsewhere\cite{RI2}, 
when a vortex-glass transition will occur within 
LLL, the presence of a gauge-invariant gradient $-{\rm i}\partial_\mu - 
(2 \pi/\phi_0) \delta A_\mu$ operating on the LLL pair-field in a 
hamiltonian such as eq.(2.4) or (3.5) implies that the corresponding 
gauge-invariant gradient operating on the glass order parameter must
appear in a resulting effective hamiltonian ${\cal H}_{{\rm eff}, \, G}$ 
expressed only by the glass order parameter, because the analysis
in LLL leading to ${\cal H}_{{\rm eff}, \, G}$ is formally the same as 
the
corresponding one in zero field case. Since this similarity to the 
zero field normal-Meissner transition implies that the scaling 
argument\cite{FFH} on the critical behavior of the conductivity 
$\sigma_{s, \mu \mu}$ near the glass transition is applicable 
just as in zero field case, we have concluded\cite{RI2} 
that the helicity modulus $\Upsilon_z({\bf q}=0)$ parallel to ${\bf 
B}$ 
should be finite in 
any vortex-glass phase. Since, in the present case, this applies 
also to the $y$-direction due to the recovery of the isotropy in
$y$-$z$ plane, we can conclude that, as in any pinned Josephson vortex
solid with no disorder\cite{RI3}, $\Upsilon_y({\bf q}=0)$ should be
also finite in the JG phase. Based on eq.(3.2), this directly implies 
that the JG phase should have a TME perpendicular to the layers, and 
consistently that the tilt response (i.e., a diamagnetic
susceptibility\cite{RI8}) $\Delta C_{44,x}$ should diverge, 
just as in the normal-Meissner transition, 
in proportional to the correlation length $\xi_{JG}(T)$ on 
approaching $T_{JG}$ from above,  $$\Delta C_{44,x} \propto \xi_{JG}(T) 
\sim (T - T_{JG}(B))^{-\nu_J}, \eqno(3.6)$$
which will reduce to eq.(3.4) far above $T_{JG}(B)$ in $p \gg 1$. 

Next, let us turn to $\Delta C_{44,y}$. In this case, it is convenient
to return to the original model (1.2) with no $\delta A_\mu$ ($\mu =
y$, 
$z$) and with replacement $\psi_j \to 
\psi_j \exp({\rm i} d 2 \pi j \, \delta A_x(z)/\phi_0)$. After 
gauge-transforming $\psi(x) \exp({\rm i} 2 \pi x \, \delta A_x(z)/\phi_0)
\to \psi(x)$ in the second representation of eq.(1.2) and, for
simplicity, assuming $\delta B_y = \partial_z \delta A_x$ to be
constant, we have 
$${\cal H}_{LD} = \int {\rm d}^3r \sum_{m = -\infty}^{+\infty} 
{\rm e}^{{\rm i} 2 
\pi m x/d} \biggl[ \varepsilon_0 |\psi({\bf r})|^2 + \xi_0^2
\sum_{\mu=y, z} \biggl|\biggl(- {\rm i}\partial_\mu + {{2 \pi} \over
{\phi_0}} (B {\hat y} - \delta B_y {\hat z}) \, x \biggr) \psi({\bf r}) 
\biggr|^2$$
$$ + \Gamma^{-1} \biggl({{\xi_0} \over d} \biggr)^2 |\psi({\bf r}) - 
\psi({\bf r} + d {\hat x})|^2 + {b \over 2} |\psi|^4 \biggr] \eqno(3.7)$$
in place of the second representation of eq.(1.2). When the disorder
term (1.3) is included in addition to eq.(3.7), the 
inclusion of $\delta B_y$ merely
implies a rotation of the applied field direction 
within $y$-$z$ plane in which the original system in zero field 
is isotropic. Namely, since this disturbance is the same as a tilt in an 
isotropic 3 D system with point disorder, no TME is expected for a tilt 
within the layers, i.e., $\Delta C_{44,y}$ is nondivergent at and below
$T_{JG}$. Note that this conclusion was obtained without using the LLL
approximation and hence, should be valid at any $p$-value. 

In contrast to this nondivergent $\Delta C_{44,y}$, the above conclusion 
on the divergent $\Delta C_{44,x}$ was obtained in terms of the high $p$ 
approximation. Hence, one may suspect that a vortex-glass phase in lower
$p$-values, where the layering is quantitatively negligible at least
above $T_m$, will have no TME as well as the
ordinary vortex-glass (VG) phase\cite{FFH}. However, as mentioned below
eq.(2.2), the cross term $\sim - \varphi^* {\rm i} \delta A_y \partial_y 
\varphi$, which determines the current vertices in the response
function, exists within LLL even without assuming a large $p$. This 
rather suggests the presence, irrespective of the $p$-value, of TME 
for a tilt perpendicular to the layers in the ordered phases. 

To clarify these points, we examine the tilt responses in the
uniaxially periodic GL model\cite{Per} with point disorder 
$${\cal H}_{\rm per} = {\cal H}_{\rm GL} - \int {\rm d}^3r [ \, 
u_p {\rm cos}(2 \pi
x/d) \, - \, u_r({\bf r}) \, ] |\psi({\bf r})|^2, \eqno(3.8)$$ 
where ${\cal H}_{\rm GL}$ is the ordinary isotropic GL model, 
the positive constant $u_p$ is the strength of a periodic
variation of $T_{c0}$ playing similar roles to the layer structure of 
the LD model, and $u_r$ is a random potential expressing the point
disorder and has the ensemble defined by ${\overline u_r}=0$ and
${\overline {u_r({\bf r}_1) \, u_r({\bf r}_2)}} = \xi_0^3 \Delta
\delta^{(3)}({\bf r}_1 - {\bf r}_2)$. 
Hereafter, we assume a weaker magnetic field satisfying $r_B
\geq d$ so that the ground state in a fixed field becomes a
$w \geq 1$ pinned solid or the floating solid.\cite{RI3,com3} Note
that, in this model, the pair-field can be nonzero not only on the
``layers'' but also between two neighboring ``layers'' and hence that
the point disorder existing between the layers is also effective as 
pinning sites enabling the vortex lines to locally deviate from
directions parallel to the layers. By contrast, in the LD model the 
point 
disorder, as well as the pair-field, is restricted on the layers 
and hence, will not play
any role for releasing the vortices locked in the layers. Due 
to this difference, described in Fig.2, in the disorder 
configurations between the two models, 
one may expect that the point defects existing {\it between} the 
neighboring ``layers'' in the periodic GL model compete with the layer 
structure inducing\cite{RI3} the nonvanishing helicity modulus
$\Upsilon_y({\bf q}=0)$, implying TME perpendicular to the layers, and 
that 
this TME perpendicular to the layers 
may be destroyed by such defects absent in the LD model. Namely, a TME 
should occur more easily in the LD model than in the periodic GL model. 
Keeping this in mind, let us examine $\Delta C_{44, \mu}$ of the model
(3.8) in a manner similar to in ref.4. 

As in previous works\cite{RI1,RI2} on glass transitions in ${\bf B}
\parallel {\bf c}$, we will use a high field approximation 
convenient in examining electro-magnetic responses. When treating
${\cal H}_{\rm GL}$ by expressing $\psi$ in terms of the Landau levels 
(LLs),
the vertex on an external current perpendicular to ${\bf B}$ is
accompanied by the next lowest LL (NLL) even if the thermodynamics is
described primarily by the LLL modes. Since this NLL mode is
not associated with any ordering in vortex systems,\cite{RI1,RI8} it
remains massive and heavy in high fields in the sense that its time
scale becomes much shorter than those of LLL modes. 
In the ideal case with no source 
of vortex pinnings, this short time scale of NLL 
inevitably becomes the time scale for all terms of the 
perturbation series of a conductivity perpendicular to ${\bf B}$, and
consequently the vortex flow conductivity follows.\cite{RI9} This
fact is clearly seen\cite{RI1} particularly 
in the high field approximation where the LLL
modes are assumed not to interact, through the $|\psi|^4$ term of 
${\cal H}_{\rm GL}$, with higher LL modes. However, when either a 
periodic or a random pinning term is present as in eq.(3.8), a much
longer time scale than that of NLL can be picked up from a section, 
composed only of the LLL modes, in Feynman diagrams 
contributing to a response quantity, and hence the presence of the 
time scale of NLL can be neglected in such diagrams. In refs.1 and 4,
such diagrams with a long time scale growing on cooling 
due to a LLL vertex correction expressing the 
VG fluctuation were found for a conductivity perpendicular to ${\bf B}$ 
in the case with no periodic 
pinning effect. Its examples are given in Fig.3 (a), where the open
square denotes the above-mentioned LLL vertex correction corresponding
to the glass susceptibility. Situation is 
similar in static responses such as $\Delta C_{44, \mu}$, and, just like
the neglect of dynamics of the NLL modes in such diagrams 
of the conductivities, the $z$-dependence of the NLL fluctuation
propagator can also be disregarded in examining $\Delta C_{44, \mu}$. 

Once the periodic potential $\propto u_p$ is included, the linear
responses become anisotropic in $x$-$y$ plane. When the gauge disturbance
$\delta {\bf A}$ couples to the pair-field through the GL hamiltonian,
the tilt response $\Delta C_{44,x}$ is generally 
expressed in the form\cite{RI1,RI2} 
$$\Delta C_{44,x} = {{2 k_{\rm B}T} \over {V \, r_B^2}} \biggl(-
{{\partial} \over {\partial k_z^2}} \biggr) \sum_{K, K'} [ 
{\overline {< \, \varphi_{1,K+k_z} \, \varphi^*_{1, K'+k_z} 
\, \varphi_{0, K'} \, \varphi^*_{0, K} \, >}} $$
$$+ \, {\overline {< \, \varphi_{1,K+k_z} \, \varphi^*_{0, K'+k_z}  
\, \varphi_{1, K'} \, \varphi^*_{0, K} \, >}} ], \eqno(3.9)$$  
where $V$ is the system volume. 
The last (second) term of eq.(3.9) vanishes 
in the case isotropic in $x$-$y$ plane. 
For the present purpose, we need not to examine eq.(3.9) in details but
only to focus on how the NLL fluctuation couples to a LLL
fluctuation, in terms of the potential terms in (3.8), outside the LLL 
vertex correction implying the glass susceptibility in Feynman 
diagrams (see Fig.3). Let us consider the diagrams in Fig.3 (b), where 
the NLL lines couple to LLL lines only through the periodic
potential, by expressing the pair-field in terms of the same
representation of LLs as in the LD model (see the sentences prior to
eq.(2.1)). The NLL-LLL coupling term, marked with dashed line in
Fig.3, is written as 
$$ - \sqrt{2} \, u_p \, {{2 \pi r_B} \over d} \exp(-(\pi r_B/d)^2) 
\sum_{Q,
q_z} {\rm sin}\biggl({{2 \pi r_B^2} \over d} q_y \biggr) \, 
( \, \varphi^*_1(n, {\bf q}_\perp) \, \varphi(n, {\bf q}_\perp) \, + \,
{\rm c. \,\, c. \,\,\,\,} ), \eqno(3.10)$$
where $\varphi_1$ is the corresponding one in NLL to $\varphi$ in 
LLL. As already stated, 
the NLL fluctuation in Fig.3 (b) can be assumed to be independent of
$z$. Then, when ensemble-averaging over the NLL modes according to Fig.3 
(c), the resulting NLL propagator can be treated as a constant 
factor multiplying 
the current vertex, and the resulting O($\delta A_y$) term ($ \propto -
u_p \, \delta A_y \, {\rm sin}(2 \pi r_B^2 q_y/d) |\varphi(n, q_y, 
z)|^2$) 
can be regarded as a current vertex defined {\it within} LLL. 
Explicitly, 
if using the fact that the NLL propagator $G_1(q_y, q_z=0) = 
< |\varphi_1(n, q_y, q_z=0)|^2 >$ in high enough fields 
can be replaced\cite{RI9}, except
corrections in $q_y^2$,  by $r_B^2/(2 \xi_0^2)$ deep in the liquid
regime,\cite{RI9} 
the obtained dispersion of the LLL fluctuation and the current vertex 
coupled to $\delta A_y$ can be seen as those defined from
the following quadratic term of an effective hamiltonian only on the LLL
modes 
$${\tilde {\cal H}}_{{\rm per}, 2} = \int {\rm d}y {\rm d}z 
\sum_n \varphi^*(n, {\bf
r}_\perp) \biggl( {u_p} \, \exp(-(\pi r_B/d)^2) \biggl( 1 - {\rm
cos}\biggl[{{2 \pi r_B^2} \over d} \biggl(- {\rm i}\partial_y - {{2 \pi} 
\over {\phi_0}} \delta A_y(z) \biggr) \biggr] \biggr)$$ 
$$ + \mu_0 + \xi_0^2 \biggl(- {\rm i} \partial_z 
- {{2 \pi} \over {\phi_0}} \delta A_z(y) \biggr)^2 \biggr) \varphi(n, 
{\bf 
r}_\perp). \eqno(3.11)$$
Now, let us consider the diagrams in Fig.3 (c) composed of the current
vertex of Fig.3 (b). In Fig.3 (c), the left (right) diagram arises from
the first (second) term of eq.(3.9). Further, note that, near $T_{JG}$,
the gauge invariant gradient term in $y$ direction has only to be kept 
up to
quadratic order. Then, the argument given below eq.(3.3) can directly be 
used in the present case. The only difference from the LD case is the
absence, at the quantitative level, of isotropy in $y$-$z$ plane due to
the difference in coefficients of the gradient terms. Since the strength 
$u_p$ is independent of other material parameters, $\Delta C_{44,x}$ in
the model (3.7) is also predicted to diverge in proportional to a
correlation length near the resulting glass transition. Contrary to
this, this divergence is absent in $\Delta C_{44,y}$, because this
quantity is given by replacing the plus sign prior to the last term 
in eq.(3.9) by a minus sign, and hence, the contributions corresponding
to the two terms of Fig.3 (c) cancel with each other. We note that the 
diagrams, such as Fig.3 (d), including just one $u_p$-vertex coupling 
between the LLL and NLL modes do not give nonvanishing contributions
because of the rule\cite{RI9} illustrated in Fig.3 (e). Hence, only the 
sum of diagrams such as Fig.3 (a) give a nonvanishing contributions 
related to
the glass fluctuation which is easily verified to be nondivergent at the 
transition for the same reason as in the VG case.\cite{RI2} In passing,
we note that what is essential to obtaining the above results is the
presence of a nonzero gauge coupling $\sim {\rm i} 
\delta A_y(z) \varphi^* \partial_y \varphi$ after averaging over the NLL 
modes but is not the use of the approximation on $G_1$ in getting 
eq.(3.11). 
 
Thus, we have shown that the divergent 
$\Delta C_{44,x}$, as well as the nondivergent $\Delta C_{44,y}$, is
also realized in this periodic GL model. We emphasize 
that this conclusion is independent of the strength of the point
disorder and of the strength $B$ of the magnetic field which
tunes the structure of the disorder-free vortex solid in ${\bf B} 
\perp {\bf c}$ through a commesurability condition\cite{RI3,IK,Per} 
with the layer structure. Namely, the glass phase in the model (3.7) 
should have the anisotropic TME {\it irrespective of} $B$. Following the
consideration based on Fig.2, this in turn convincingly suggests 
that our earlier prediction on the LD model 
that the resulting JG phase has TME only for tilts perpendicular to the
layers is also likely to be valid for any $B$. Namely, the ground state
will become the JG phase with a TME even in $p$ ($< 1$)-values where the
{\it disorder-free} ground state may be a unpinned (floating) vortex 
solid, 
as far as the positional order in the resulting JG phase at such
$p$-values is short-ranged. Further, we note that the presence of the 
TME perpendicular to the layers below $T_{JG}$ for any $p$ implies the 
presence in any $B$ of the so-called lock-in transition due to a tilting
of the applied field in real systems {\it with} 
point disorder. Note that the so-called lock-in transition close to
${\bf B} \perp {\bf c}$ was discussed so far just in the pinning-free
case. Since the floating solid possible in the
disorder-free case has no TME, the above result implies that the point
disorder destroying the positional order of this solid simultaneously
creates the anisotropic TME in the real JG phase. 

\section{${\bf B} \perp {\bf c}$ Phase diagram and Related Issues}

The resulting ${\bf B} \perp {\bf c}$ phase diagrams in this paper are 
sketched in Fig.4. The main point in our 
results is that the resulting $T_{JG}$ must decrease with increasing $B$
even in high fields satisfying $B > \phi_0/\sqrt{\Gamma} d^2$, while the 
first order melting temperature $T_m$ and phenomena in the vortex liquid 
regime above $T_m(B)$ in such high fields will be almost 
independent\cite{RI3} of $p \propto B \sqrt{\Gamma}$, reflecting the 
nearly complete confinement of vortices 
within the interlayer spacings. The origin of the field dependence of
$T_{JG}$ is a spatial variation of $|\psi|$ on the superconducting 
layers, 
which is small but nonvanishing in $B > \phi_0/(\sqrt{\Gamma} d^2)$, 
and, as already mentioned, is indispensable to a glass ordering. 
Since this spatial variation strongly depends on $p$ 
and disappears in high $B$ (or high $\Gamma$) {\it limit} of 
the LD model, $T_{JG}$ at which the 
resistivities vanish {\it must} depend on both $B$ and $\Gamma$ and 
approach zero in this limit. Namely, in ${\bf B} \perp {\bf c}$ case, 
$T_{JG}$ rapidly starts to deviate 
from the (disorder-free) melting line $T_m(B)$ with increasing $B$ 
even in clean systems, and hence, the resistive curves just above
$T_{JG}$ but below $T_m$ 
also become $B$-dependent. This is quite different from the (ordinary) 
VG transition line\cite{RI1,FFH} $T_{VG}(B)$ in 
${\bf B} \parallel {\bf c}$ case, in 
which the VG line $T_{VG}(B)$ in higher fields (described as the lower 
dashed 
curves in Fig.7 of ref.1) yields the LLL scaling\cite{RI9} similar to
that of $T_m(B)$, and for this reason, this $T_{VG}(B)$ line is often 
regarded, for a practical purpose, as an extrapolation 
of the $T_m(B)$ line by neglecting a narrow window\cite{RI1} 
of the slush regime observed\cite{Koban} recently in YBCO. Therefore, 
particularly in materials with stronger fluctuation, 
there may be a much wider vortex 
slush\cite{RI1,Wor} regime in ${\bf B} \perp {\bf c}$ than in ${\bf B} 
\parallel {\bf c}$. Further, as shown in $\S 3$, the ``force free''
behavior itself in such high fields is never affected by the point
disorder isotropic on the superconducting layers. Consequently, the 
in-plane resistivity curves in $p > 1$ will show, near $T_m$, a
crossover from a $B$-independent ``force free'' behavior to a
$B$-dependent ``force free'' behavior below $T_m$. 
The $B$-{\it dependent} ``force-free'' resistive 
behavior below $T_m$ is qualitatively consistent with the behavior 
at {\it lower} temperatures seen in the data of ref.11. 

For convenience, we have distinguished a dirtier case Fig.4 (a) from a
cleaner case Fig.4 (b) based on whether a critical point of the first
order transition line, denoted as $p=p_c$, satisfies $p_c < 1$ or $p_c 
> 
1$. First, let us consider Fig.4(a). In this case, with 
increasing $p$, $N_c$ may already become of order unity before entering
the $p > 1$ region (Note that the shortest value of the 
positional correlation length perpendicular to the layers is the layer
spacing $d$ and hence that one can say $N_c|_{\rm min} \sim$ O($1$)). 
Then, the relation 
$B_c^{({\rm in})} \ll \phi_0/(2 \pi d^2 \sqrt{\Gamma} \, 
\varepsilon_G^{(2)})$ 
will be satisfied over a wide range 
in $p > 1$, and we will have the relation $T_{c2} - T_{JG} \propto
T_{JG} \, B$ (see eq.(2.24)). Next, in turn, let us assume 
a cleaner case with $p_c > 1$ such as Fig.4 (b). 
In this case, $N_c$ in $p < p_c$ will be of order $N_d$. As already 
mentioned, this means that the system below $p_c$ is in clean limit and 
that $T_{JG} \simeq T_m$ there (see Fig.4(b)). Next, when $p > p_c$, 
$N_c$ will rapidly decrease with 
increasing $p$, because the correlation volume will decrease
with decreasing the shear modulus\cite{RI3,IK} $\sim {\rm e}^{-p}$ in 
the $w=1$ solid-like state. If naively assuming $N_c \sim {\rm 
e}^{-c_1 p}$ 
with a constant $c_1 > 0$, the neglect in eq.(2.22) of a $B$-dependence 
of
$B_c^{({\rm in})}$ is not valid any longer. Only in $p$ values larger 
than 
$p_{c1}$, $N_c$ will start saturating into a constant of order unity, 
and hence $B_c^{({\rm in})}$ approaches a
small positive or a negative constant. Namely, eq.(2.24) becomes valid
only when $p > p_{c1}$. Consequently, we have 
conjectured the phase transition curves, the solid curves in Fig.4 (a)
and (b). It should be emphasized 
that the slope of the behavior $B_{JG} \sim T_{c0} - T$ to be seen 
below $T_m$ is gentler as the fluctuation is 
stronger and is {\it insensitive} to the pinning strength. 
This will become useful for examining $T_{JG}(B)$ experimentally and
implies that the vortex slush region\cite{RI1,Wor} to be seen
experimentally becomes 
wider in materials with stronger fluctuation. According to ref.1, such
an extension of the temperature window of the vortex slush region 
due to an enhanced fluctuation should also occur in 
${\bf B} \parallel {\bf c}$ case\cite{RI1}, contrary to the Bragg glass 
scenario\cite{GD2} of the ${\bf B} \parallel {\bf c}$ phase diagram, 
and, in fact, was recently observed\cite{Koban} through a doping
dependence of the $B$-$T$ phase diagram of YBCO.  

Theoretically, studies of a glass phase in ${\bf B} \perp {\bf c}$ were
performed in previous works\cite{GD,Natter} by assuming a
dislocation-free glass phase. However, it was argued\cite{Natter2} that 
the resulting transition is not accompanied by a divergence of a 
glass correlation length, just like a vortex liquid-Bragg glass 
transition argued as a scenario of the first order transition 
in ${\bf B} \parallel {\bf c}$ case. Further, since 
the assumption in refs.2 and 3 that the vortices cannot move across 
the layers is valid only in\cite{RI3} $p > 1$, the situation assumed in 
these works\cite{GD,Natter} is, at most, limited to the narrow region 
$1 < p < p_c$ of Fig.4 (b). 

As shown in ref.5, the positional correlation of vortices perpendicular 
to
${\bf B}$ in ${\bf B} \perp {\bf c}$ first develops, on cooling, 
in $y$ direction {\it parallel} to the layers. This is a consequence of 
the recovery of isotropy in $y$-$z$ plane in high enough fields, 
emphasized in $\S 2$, which is essential to understanding the 
force-free behavior\cite{Kad,Iye2} of the in-plane resistivities. 
On the other hand, the above-mentioned feature that the positional
ordering first grows not across but rather along the layers contradicts
the conjecture\cite{BN} favoring a smectic liquid phase. Namely, the
smectic liquid picture\cite{BN} is incompatible with the force free
resistive behavior. 

In ref.38, Hu and Tachiki still argue that the {\it disorder-free}
superconducting transition in ${\bf B} \perp {\bf c}$ is continuous for
cases with larger anisotropy in which the anisotropy-induced 
vortex loop excitations seem to play an important role in the
transition, while it is reasonably of first order for smaller
anisotropies (see Note added in proof of ref.5). It is remarkable that 
the transition temperature they found is insensitive, at a fixed field,
to the anisotropy even in $1.5 < p < 2$ studied in ref.38 (see Fig.3
there-in). This is consistent with our prediction\cite{RI3} that, in $p
> 1$, $T_m(p)$ approaches $T_{sc}$ independent of $p \propto B
\sqrt{\Gamma}$, because, as commonly seen in various 
works\cite{Blatter,KL,IK} on the phase-only model, the magnetic 
field $B$ always appears only as the parameter $p$. Hence, the
transition temperature in ref.38 is identified with $T_m \simeq
T_{sc}$. Clearly, this transition temperature has nothing to do with 
the observed\cite{Ishiguro,Kad2} 
superconducting transition temperature in BSCCO which significantly
decreases with increasing $B$ and also depends\cite{Ishiguro} much on 
the 
doping level, i.e., on the anisotropy (see below). 
 
After preparing the first draft of the present paper, we were aware of 
recent data of a.c. susceptibility\cite{Ishiguro} and the 
resistivities\cite{Kad2,Gri}, both of which can be direct probes of a 
glass transition, in field configurations including ${\bf B} \perp {\bf 
c}$ case of clean crystals of high $T_c$ superconductors. 
We will not discuss here complicated
behaviors\cite{Kad2} in fields tilted from the layers but focus on the
data in ${\bf B} \perp {\bf c}$ relevant to the present work. In ref.19, 
the superconducting transition field in ${\bf B} \perp {\bf c}$ was
determined in BSCCO crystals 
as the position at which the lock-in behavior (i.e., the TME 
in ${\bf B} \perp {\bf c}$ studied in $\S 3$) seen in 
a.c. susceptibility data disappears with increasing the in-plane field.  
 
The resulting line $H_\parallel^{cr}(T)$ roughly 
obeys $\Gamma H_\parallel^{cr} \sim (T_{c0} - T)/T$, which is the same
one as eq.(2.24) or (2.27) if the prefactor $\Gamma$ is replaced by
$\sqrt{\Gamma} \, \varepsilon_G^{(2)}$. However, this difference in the
prefactor does not 
preclude this identification between $H_\parallel^{cr}$ and $B_{JG}$ 
because the fluctuation strength $\varepsilon_G^{(2)}$ proportional 
to the squared penetration depth also increases, as well as the 
anisotropy $\Gamma$, with decreasing the doping level. 
Using a resonable $\varepsilon_G^{(2)}$ value in under-doped BSCCO, 
$\varepsilon_G^{(2)} \simeq 0.2$, corresponding to the in-plane 
penetration depth $\lambda(T=0) \sim 3000 (A)$, the 
$H_\parallel^{cr}(T)$-line in the underdoped case of ref.19 is 
quantitatively consistent with eq.(2.27), i.e., $B_{JG}(T)$ obtained 
by assuming $B_c^{(in)} \ll \phi_0/(2 \pi d^2 \sqrt{\Gamma} \, 
\varepsilon_G^{(2)}) \simeq 3$ (T). On the other hand, the corresponding 
transition fields in BSCCO were determined in ref.39 as the positions at 
which
the in-plane resistance at $H_c = 0$ vanishes. According to ref.39, 
this 
transition is continuous. The resulting transition field in an 
optimally-doped sample seems to be comparable with the data in ref.19
and is roughly linear in temperature just like eq.(2.27). Further, in 
resistive measurements in YBCO\cite{Gri}, resistivities in all
directions were found to vanish at the same temperature $T_i$. This is
an evidence of a continuous JG transition occuring below the
disorder-free melting line (see $\S 1$). It seems to us that, in ref.40, 
the
out-of-plane resistivity ($\rho_c$) vanishes more rapidly (with a
smaller exponent) compared with the in-plane 
resistivities. Explaining this behavior seems to be theoretically 
interesting and will be tried elsewhere. 

\vspace{5mm}

\leftline{\bf Acknowledgement}

We gratefully acknowledge an informative discussion with T. Ishiguro and 
S. Nakaharai. This work was supported by a grant for CREST from Japan
Science and Technology Corporation.

\begin{figure}
\begin{center}
 \leavevmode
 \epsfysize=7.5cm
 \epsfbox{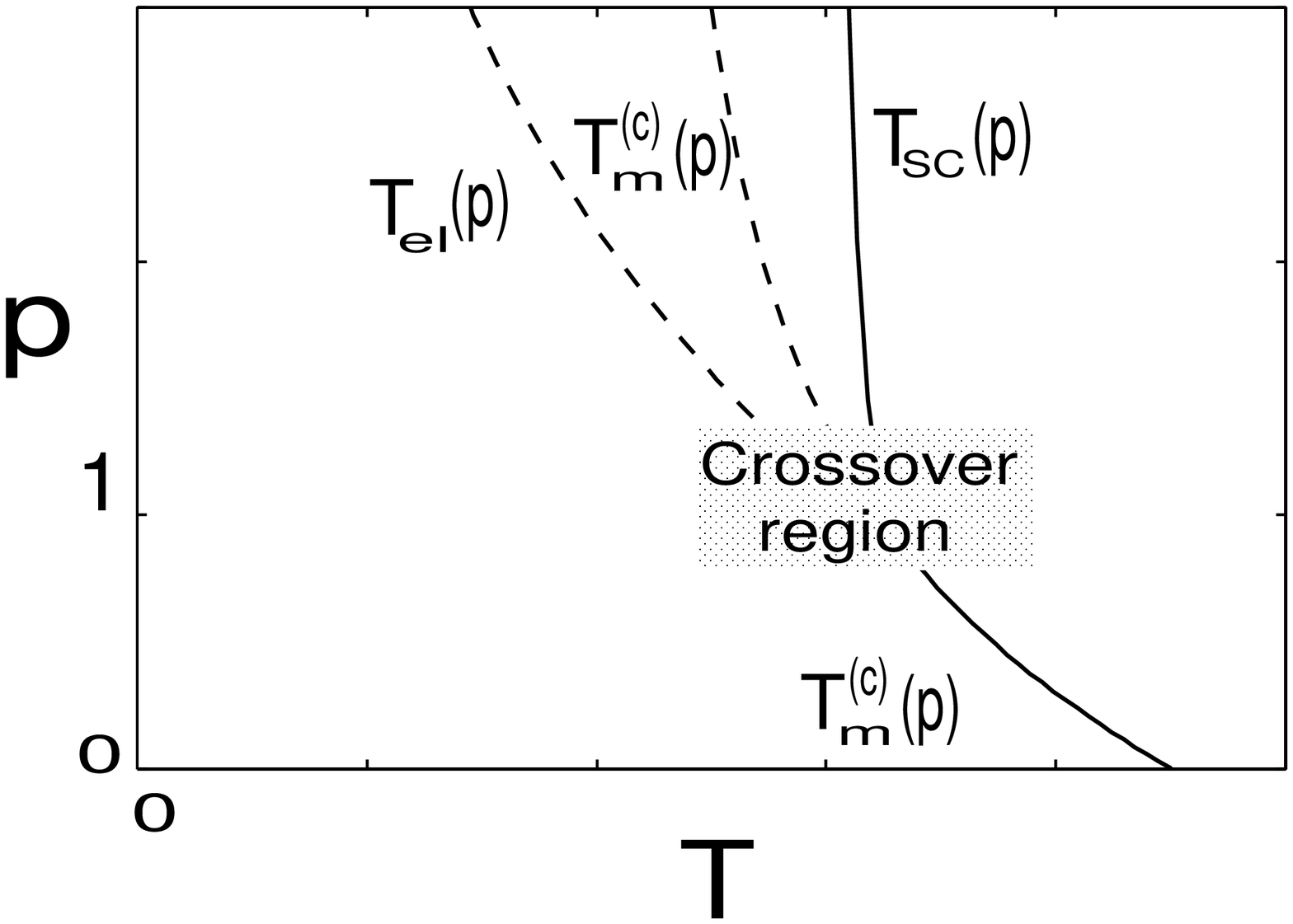}
\end{center}
 \caption{Three possible characteristic curves, $T_{el}(p)$,
 $T_m^{(c)}(p)$, and $T_{sc}$, in the $p$ v.s. $T$ diagram of the
 phase-only approximation of the LD model. These curves merge roughly
 when $p \simeq 1$. The solid curve is believed to be the true
 transition curve in the disorder-free case. }
\end{figure}

\begin{figure}
\begin{center}
 \leavevmode
 \epsfysize=7.5cm
 \epsfbox{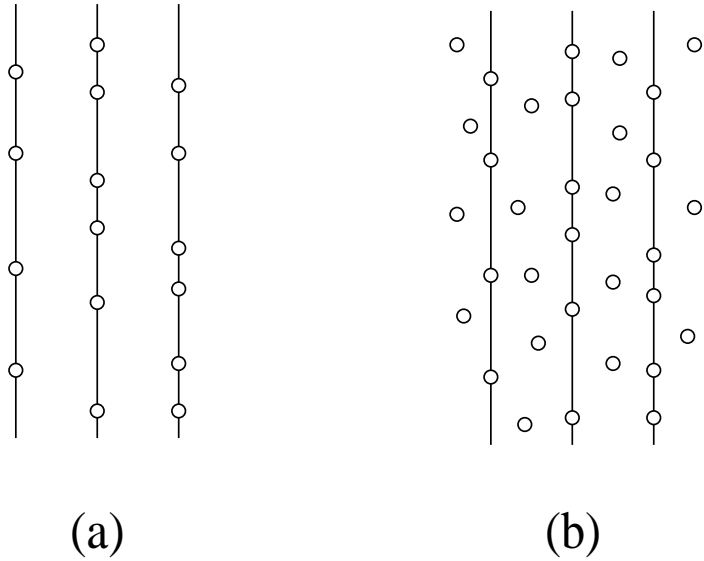}
\end{center}
 \caption{Configurations of point disorder in (a) LD and (b) the
 periodic GL models. Figures are described in $x$-$y$ plane. Solid
 lines, in $x$ direction, denote the ``superconducting layers'', and the
 open circles denote the point defects for the pair-field.  }
\end{figure}

\begin{figure}
\begin{center}
 \leavevmode
 \epsfxsize=12cm
 \epsfbox{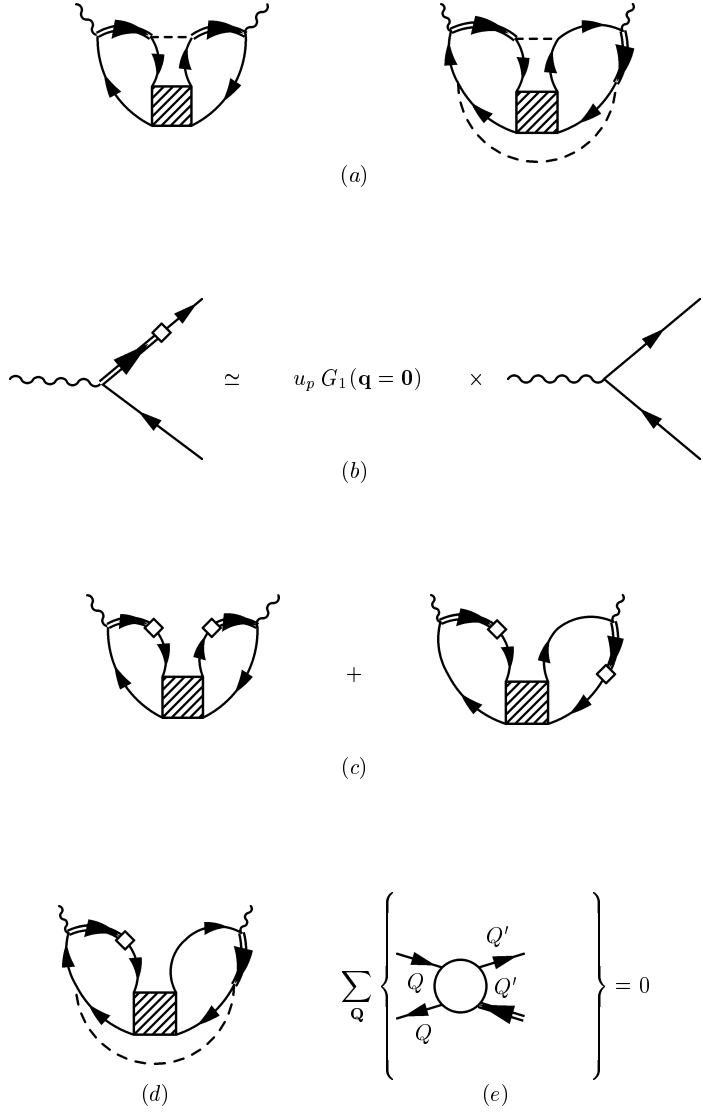}
\end{center}
\caption{Feynman diagrams on tilt moduli in the high field
 approximation\cite{RI1}. The dashed line denotes the impurity line 
 occuring after averaging over the point defect configurations and
 carries the factor $\Delta$, and the connection marked by an open
 square between a NLL propagator $G_1({\bf q})$ (the double-solid line
 with arrow) and a LLL propagator (the solid line with arrow) implies the
 periodic potential with strength $u_p$. The hatched square and open
 sphere in (e) imply four-point vertices composed only of the LLL modes.
 See the text for further details. } 
\end{figure}

\begin{figure}
\begin{center}
 \leavevmode
 \epsfysize=18cm
 \epsfbox{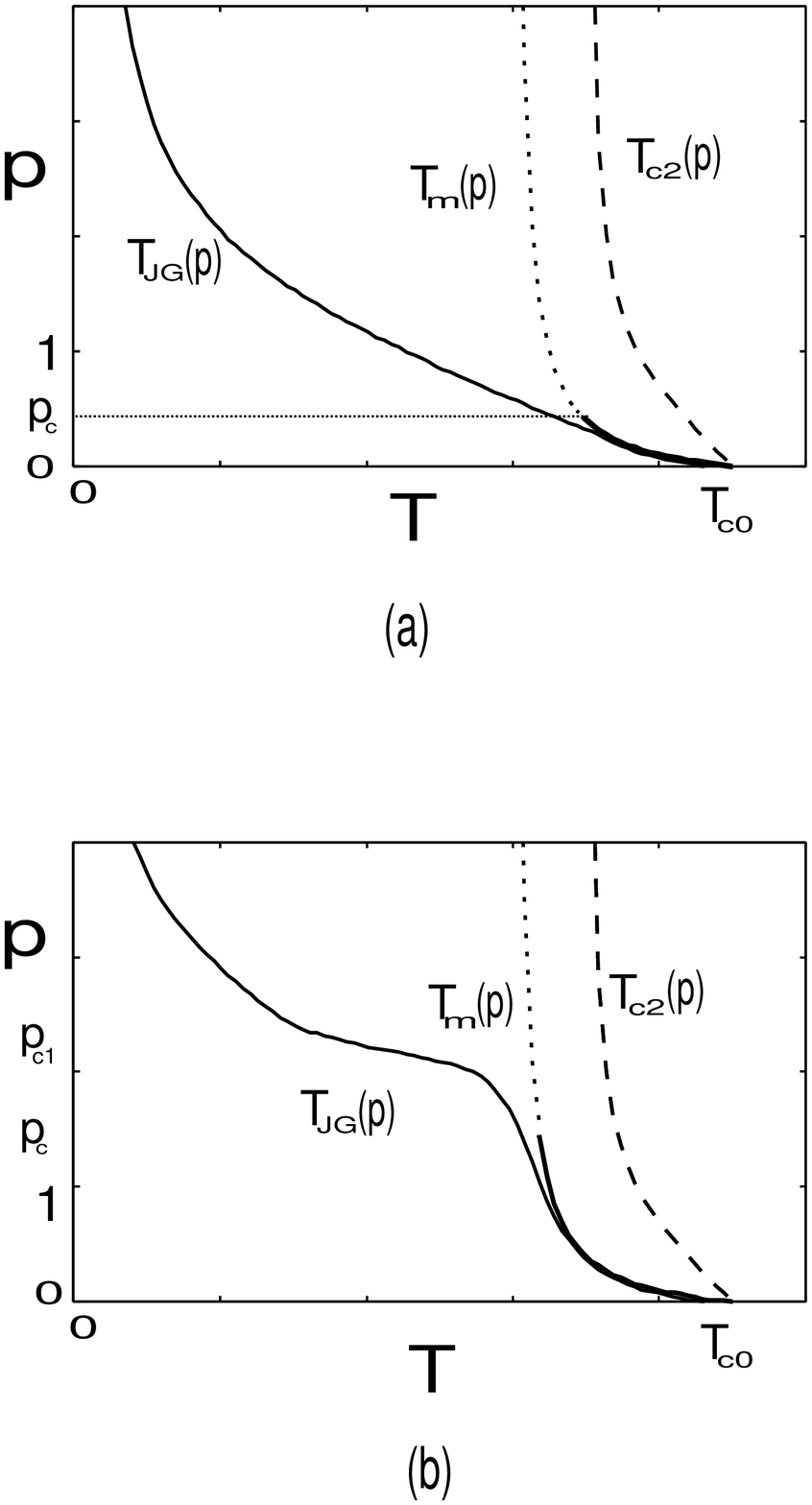}
\end{center}
\caption{Schematic $p$ v.s. $T$ phase diagrams of real layered
 superconductors in ${\bf B} \parallel$ layers for (a) dirtier and (b)
 cleaner cases resulting from the present study. The value $p_c$, the
 $p$ value at which the first order transition (expressed by the solid
 curve on $T_m(p)$) ends, measures the disorder strength. The relation
 (2.24) will be satisfied entirely in $p > p_c$ in the dirtier case (a),
 while it is valid only in $p > p_{c1} (> p_c)$ in case (b). In each
 figure, the left solid curve $T_{JG}(p)$ is the JG transition line,
 while the right dashed curve $T_{c2}(p)$ indicates the $H_{c2}(T)$
 curve. }  
\end{figure}


\begin{thebibliography}{99}
\bibitem{com1} Further, note also that the quantity $\beta_A - 1$
	relevant to the positional ordering becomes significantly
	smaller with increasing $p$ ($> 1$) due to the layering (or, an
	effect of the uniaxial periodic potential). It suggests that the 
	jump of physical quantities at the solidification transition 
is weakened with increasing $p$. At a fixed magnitude of $\Delta_{\rm 
eff}$, 
a weaker first order transition will be more easily destroyed by this
	pinning disorder.
\end{thebibliography}
\end{document}